\documentclass[aps,prd,onecolumn,groupedaddress]{revtex4}

\usepackage{epsf,epsfig,rotate}
\usepackage{t1enc}
\usepackage{amsmath,amsthm,amsfonts,amssymb,wrapfig}
\usepackage{colordvi}

\begin{document}  


\title{Constraints on the size of the extra dimension from Kaluza-Klein gravitino decay}
\author{David Gherson}
\affiliation {Institut de Physique Nucl\'eaire de Lyon (IPNL), Universit\'e Lyon-I, Villeurbanne, France}

\date{\today}

\begin{abstract}

We study the consequences of the gravitino decay into dark matter. We suppose that the lightest neutralino is the main component of dark matter. In our framework gravitino is heavy enough to decay before Big Bang Nucleosynthesis starts. We consider a model coming from a five dimensional supergravity compactified on $S^1/Z_2$ with gravity in the bulk and matter localized on tensionless branes at the orbifold fixed points. We require that the dark matter, which is produced thermally and in the decay of Kaluza-Klein modes of gravitino, has an abundance compatible with observation. We deduce from our model that there are curves of constraints between the size of the extra-dimension and the reheating temperature of the universe after inflation. 

\end{abstract}

\pacs{}

\maketitle

\section{Introduction} 			      

The five dimensional picture of the Universe has attracted much interest in the framework of what is called brane world cosmology. Several models have been proposed. It has been shown \cite{Csaki:1999mp} that a stabilization of the radion field provides the classical Friedman equation. Many papers (see for instance: \cite{Falkowski:2001sq}) have considered a supersymmetric version of brane world which embed brane world in the framework of superstrings theories. In the present work we choose to work in a set-up where we assume that the radion is stabilized in a five dimensional supergravity compactified on $S^1/Z_2$ where matter and gauge fields live on the branes and gravity in the bulk (\cite{Bagger:2001ep}, \cite{DeCurtis:2003hs}). Low energy supersymmetry provides a natural candidate for dark matter if R-Parity \cite{Barbier:2004ez} is conserved and solves the hierarchy problem. We have to point out that in the present model the extra-dimension does not play a role to solve the hierarchy problem like in ADD model (\cite{Arkani-Hamed:1998rs}, \cite{Antoniadis:1998ig}) or Randall Sundrum \cite{Randall:1999ee}. It is supersymmetry which plays this role. So the size of this extra dimension is not constrained like in \cite{Hannestad:2001nq} and actually we are dealing with smaller extra-dimension than in those scenarios.\\ 
As a general framework we can choose a scenario in which susy breaking is mediated to the observable sector partly by gravity anomaly and partly by Scherk-Schwarz mechanism \cite{Rattazzi:2003rj}. It is a scenario which can avoid the appearance of tachionic masses which are present in pure anomaly mediated scenario and problems due to pure gravity mediation scenario. 
This mechanism provides high masses for gravitino (it means that the mass of the gravitino can be above $10$ TeV).\\
The gravitino field has Kaluza-Klein excitations modes due to the presence of the extra-dimension. We suppose that all gravitino modes are produced after inflation during the reheating period by scattering effects in the primordial thermal bath. 
We suppose that the lightest mode - i.e. the zero mode - is heavy enough to mainly decay before the Big Bang Nucleosynthesis starts and we calculate its corresponding mass. Actually this is naturally the case in some scenarios of susy breaking (anomaly mediation or mix between anomaly and Scherk-Schwarz mechanism). Gravitinos modes decay into supersymmetric particles and standard model ones. If R-parity is conserved, all gravitino modes will give at the end of their decay cascade a Lightest Supersymmetric Particle (LSP) which is assumed to be the lightest neutralino. But not all of these decay products of gravitino will contribute to the relic density of neutralinos which is assumed to be the dark matter. Actually only the gravitino modes which decay after the thermal decoupling of neutralino contribute to the dark matter density. If a gravitino mode decays before the thermal decoupling of neutralinos it does not increase the number density of neutralinos since these gravitinos produce neutralinos which are in thermal equilibrium. So a finite number of gravitinos modes contribute to the non thermally produced dark matter. The total of the thermally produced and non thermally produced amount of dark matter is constrained by the evaluation of the dark matter content of the universe. As a consequence, we can draw curves of constraints between the size of the extra-dimension and the reheating temperature because the number of KK gravitinos modes is related to the size of the extra-dimension and the number density of gravitinos is related to the reheating temperature. We have chosen a reheating temperature in the range $10^5$ GeV to $10^{10}$ GeV. This range is quite natural for scenarios which allow baryogenesis through leptogenesis. There is another constraint on the size of the extra-dimension coming from the fact that there are also KK gravitons which can disturb BBN if the number of KK modes is too high.\\ 
In the present work we first present the interactions between KK gravitino and MSSM, and discuss abundance and lifetime of gravitino. After this, we estimate the thermal production of neutralinos function of $x_f=\frac{m_{lsp}}{T_f}$ where $T_f$ is the neutralinos freezing temperature and $m_{lsp}$ the neutralino mass. We then present the model, the equations of constraints and check that KK gravitons are not a problem for BBN for $R^{-1}>1$ TeV. Finally we show and comment the results.

\section{Interactions between gravitino and MSSM}

The four dimensional (reduced) Planck mass M is related to the five dimensional one $M_5$ by the relation $M^2=\pi\,R\,M_5^3$ where $R=\frac{r}{M_5}$ is the physical radius, $r$ is an undetermined real constant and $M_5=\kappa^{-1}$.\\ 
As matter and gauge fields live on the brane, those fields do not have a dependence on the fifth coordinate $x^5$. The two gravitino fields (odd and even under $Z_2$) have the following Fourier expansion:\\
for the even fields under $Z_2$:
\begin{equation}
\psi^{\rm even}_\mu(x^\lambda,x^5) =  {1 \over \sqrt{\pi r}} \left[ \psi^{\rm
even}_{0,\mu}(x^{\lambda}) + \sqrt{2} \sum_{n=1}^{\infty} \psi^{\rm even}_{n,\mu}(x^{\lambda})
\cos(n M_5 x^5) \right] \, , \label{grava}
\end{equation}
and the odd fields:
\begin{equation}
\psi^{\rm odd}_\mu(x^\lambda,x^5)={\sqrt{2} \over \sqrt{\pi r}}
 \sum_{n=1}^{\infty} \psi^{\rm odd}_{n,\mu}(x^{\lambda}) \sin(n M_5 x^5)\, ,
\label{grav}
\end{equation}

We can define:

\begin{equation}
\psi_{\mu}(x^{\lambda},x^5)=\psi^{\rm even}_\mu(x^\lambda,x^5)+\psi^{\rm odd}_\mu(x^\lambda,x^5)
\label{gravb}
\end{equation}

$\psi^{\rm even}_{n,\mu}(x^{\lambda})$ and $\psi^{\rm odd}_{n,\mu}(x^{\lambda})$ are the Kaluza-Klein modes.\\

Here are the relevant terms of interactions between matter, gauge fields and gravitino in the four dimensional supergravity lagrangian:

\begin{eqnarray}
{\cal L}_{inter}^{4d}  =
-\frac{1}{\sqrt{2}M} e g_{ij^*} \tilde{\cal D}_{\nu} \phi^{*j}
\chi^i \sigma^\mu \bar{\sigma}^\nu \psi_\mu
-\frac{1}{\sqrt{2}M} e g_{ij^*} \tilde{\cal D}_{\nu} \phi^{i}
\overline{\chi}^j \bar{\sigma}^\mu \sigma^\nu \overline{\psi}_\mu
\nonumber          \\ 
-\frac{i}{2M} e \left(
\psi_\mu \sigma^{\nu\rho} \sigma^\mu \overline{\lambda}_{(a)}
+ \overline{\psi}_\mu \bar{\sigma}^{\nu\rho}
\bar{\sigma}^\mu \lambda_{(a)} \right)
F_{\nu\rho}^{(a)},
\label{L4}
\end{eqnarray}

Where $\phi$ are scalar fields, $\chi$ are chiral fermions, $\lambda$ are gauge fermions (gauginos) , $F_{\mu\nu}^{(a)}$ is the field strength tensor for the gauge boson $ A_\mu^{(a)}$. Indices i, j..represent species of chiral multiplets and (a), (b).... are indices for adjoint representation of gauge group. e is the vierbein. $ g_{ij^*}$ is the Kahler metric.\\
This lagrangian (\ref{L4}) describes the interaction of the $0$ mode with matter and gauge fields.\\ 
We obtain the four dimensional lagrangian by integrating over the fifth dimension the five dimensional one. As the gauge fields and matter fields of the observable sector live on one brane (for instance the one located at $x^5=0$), the part of the action which contains interaction of the gravitino field  with the matter and gauge fields is  : 
  
\begin{eqnarray}
S = \kappa\int d^4x \int_{-\pi\kappa}^{\pi\kappa}dx^5\, \delta(x^5)
( -\frac{1}{\sqrt{2}} e g_{ij^*} \tilde{\cal D}_{\nu} \phi^{*j}
\chi^i \sigma^\mu \bar{\sigma}^\nu \psi_\mu(x^\lambda\,,x^5)
-\frac{1}{\sqrt{2}} e g_{ij^*} \tilde{\cal D}_{\nu} \phi^{i}
\overline{\chi}^j \bar{\sigma}^\mu \sigma^\nu \overline{\psi}_\mu(x^\lambda\,,x^5)
\nonumber          \\ 
-\frac{i}{2} e \left(
\psi_\mu(x^\lambda\,,x^5) \sigma^{\nu\rho} \sigma^\mu \overline{\lambda}_{(a)}
+ \overline{\psi}_\mu(x^\lambda\,,x^5) \bar{\sigma}^{\nu\rho}
\bar{\sigma}^\mu \lambda_{(a)}\right)
F_{\nu\rho}^{(a)}) 
\label{L5}
\end{eqnarray}

Using the definition (\ref{gravb}) and inserting the Fourier expansion of gravitino (\ref{grava}) and (\ref{grav}) into the above action, we find the interactions of each KK mode with matter and gauge fields after integrating on the fifth dimension. After redefining KK modes of gravitino (\cite{Bagger:2001ep}, \cite{DeCurtis:2003hs}):
\begin{equation}
\psi_{n,\mu}=\frac{\psi^{\rm even}_{n,\mu}+ \psi^{\rm odd}_{n,\mu}}{\sqrt{2}},\ {\rm for} \ n>0
\label{redef}
\end{equation}

and 

\begin{equation}
\psi_{0,\mu}=\psi^{\rm even}_{0,\mu}
\end{equation}

we find:

\begin{eqnarray}
{\cal L}_{interKK}^{4d}=\sum_{n=0}^{\infty} 
( -\frac{1}{\sqrt{2}M} e g_{ij^*} \tilde{\cal D}_{\nu} \phi^{*j}
\chi^i \sigma^\mu \bar{\sigma}^\nu \psi_{n,\mu}
-\frac{1}{\sqrt{2}M} e g_{ij^*} \tilde{\cal D}_{\nu} \phi^{i}
\overline{\chi}^j \bar{\sigma}^\mu \sigma^\nu \overline{\psi}_{n,\mu}
\nonumber          \\ 
-\frac{i}{2M} e \left(
\psi_{n,\mu} \sigma^{\nu\lambda} \sigma^\mu \overline{\lambda}_{(a)}
+ \overline{\psi}_{n,\mu} \bar{\sigma}^{\nu\lambda}
\bar{\sigma}^\mu \lambda_{(a)} \right)
F_{\nu\lambda}^{(a)})
\label{L4new}
\end{eqnarray}

So we get an infinite sum of the Lagrangian (\ref{L4}). Each Kaluza-Klein mode has the same interaction with matter and gauge fields. This is not a surprise in the sense that graviton and gravitino are in the same supermultiplet and each KK gravitons has the same interaction with matter and gauge fields \cite{Han:1998sg}.\\
Gravitino Kaluza-Klein modes differ only by their masses. They get a mass through the super-higgs mechanism. \\
Other redefinitions of the gravitinos fields have to be performed to obtain the gravitino mass matrix. Those redefinitions do not play a role in the interaction part. After diagonalization of the mass matrix there are no interactions between the different KK modes.\\
The masses of the modes n are related to the mass of the zero mode by the relation (\cite{Bagger:2001ep},\cite{DeCurtis:2003hs}):

\begin{equation}
M_{n}=M_0 + \frac{n}{R}
\label{mass}
\end{equation}

\section{Abundances}

 In the present work we suppose\footnote{There is a possibility of the gravitino production from the decay of scalar particles \cite{Endo:2006tf} and \cite{Endo:2006zj}. In our case the mass of the radion that we call $\Phi$ is very small \cite{Antoniadis:1998ig} of the order of $10^{-4}$ eV and so it cannot decay into the gravitinos. But with such a mass the radion is a problem for cosmology  because it can dominate the universe energy density if the value of its potential at $\Phi=0$ is close to the Planck mass \cite{Ellis:1987qs}. This problem is evaded if the value of the potential at $\Phi=0 $ is between $10^{10}$ and $10^{12}$ GeV see \cite{Ellis:1987qs}.} that gravitinos are produced during the reheating period after inflation by scattering processes\footnote{A group \cite{Rychkov:2007uq} has very recently done the calculation adding production via decay and effect of the top Yukawa coupling.}. All the MSSM particles are in thermal equilibrium except the gravitino which decouples as soon as it is produced because of its extremely weak interaction (gravitational one). Ten reactions are considered to calculate the gravitino production. Infrared divergences appear in the calculation due to the exchange of massless gauge bosons in the t-channel. This problem is avoided by the method developed in \cite{Bolz:2000fu}. Another group (\cite{Pradler:2006qh},\cite{Pradler:2006hh}) has recently done the calculation again with a slightly difference in the result. Abundance is given in \cite{Kohri:2005wn} taking into account production during the inflaton-dominated epoch and \cite{Pradler:2006hh} has the same result without taking into account production during inflation. This result is given for masses of gravitino much higher than gauginos masses but the calculation of the creation term in the Boltzmann equation is made with particles without masses: their mass is supposed negligible compared to the average energy in the center of mass of each reaction. The formula is :

\begin{eqnarray}
      Y_{3/2} &\simeq& 
    1.9 \times 10^{-12}
    \nonumber \\ &&
    \times \left( \frac{T_{\rm R}}{10^{10}\ {\rm GeV}} \right)
    \left[ 1 
        + 0.045 \ln \left( \frac{T_{\rm R}}{10^{10}\ {\rm GeV}} 
        \right) \right]
    \left[ 1 
        - 0.028 \ln \left( \frac{T_{\rm R}}{10^{10}\ {\rm GeV} } 
        \right) \right],
    \nonumber \\
\label{Ygrav}
\end{eqnarray}

where $Y_{3/2}=\frac{n_{3/2}}{s}$, $n_{3/2}$ is the number density, $s$ is the entropy density and $T_R$ the reheating temperature.
The quantity $Y=\frac{n}{s}$ is the density per comoving volume.\\

We have to take into account in our calculation the different masses of gravitino modes. We find as a good approximation this rule for the abundance of the different modes:

\begin{eqnarray}
Y_{3/2}^k = Y_{3/2}^0 \ ,& {\rm for}\  M^k\le T_{\rm R}& \rm{and}\nonumber\\
Y_{3/2}^k = 0\ ,& {\rm for}\ M^k > T_{\rm R}
\label{abondancekk}
\end{eqnarray}

where k represents the index of the KK mode, $M^k$ is the mass of the $k^{th}$ gravitino mode and $Y_{3/2}^k$ its abundance. \\
In this model all particles are considered massless compared to the reheating temperature scale except the gravitino. All particles in the thermal bath are relativistic at $T_R$ except the gravitino modes which can be non relativistic if heavy enough. 

\section{Decay of the gravitino modes}

In a previous section we showed that each gravitino mode had the same interaction with matter and gauge fields. The only difference arises from the mass of the different modes.\\
Gravitino interacts gravitationally with MSSM particles: its life time is long compared to the one of other particles. We have calculated the life time for heavy gravitino (${\rm masses} > 10\ {\rm TeV}$) using \cite{Kohri:2005wn}:

\begin{eqnarray}
\tau_k = 1.4\ 10^7\times \left( \frac{M_k}{100 {\rm GeV}} \right)^{-3} \ {\rm Sec}
\label{tempsdevie}
\end{eqnarray}

In our model R-parity is conserved and the LSP is the lightest neutralino.\\
Several decay channels are possible. Heavy gravitino decays mainly into squark-quark pairs or gluon-gluino pairs \cite{Kohri:2005wn}. We checked that finally one gravitino produces one non-thermal neutralino and not less because number density of susy particles produced by gravitino decay is so weak ($Y \approx 10^{-12}$) that they will more likely decay than interact to produce standard model particles: in the Boltzmann equation the decay term is proportional to $Y$ and the annihilation term is proportional to $Y^2$. We also checked that annihilation of the LSP can be regarded as negligible because it will concern only the LSP produced by one gravitino mode when the total of modes is three. Gravitino does not produce more than one LSP because even high energy quark or gluon produced by gravitino decay if they interact between them or with a quark or a gluon of the plasma will almost all time produce quarks or gluons  rather than neutralinos as processes with strong interactions are favored compared to the ones with weak interactions. We also checked that the decay of gravitino modes does not increase entropy density because the energy density of each gravitino mode at the time of its decay is never higher than the radiation energy density at the same time: it is because all the gravitino modes decay before $T=1$ MeV. \\

\section{Neutralinos} 
In our model the LSP is the lightest neutralino. We choose to work with a mass of LSP equal to $120$ GeV. This analysis can be easily rescaled for another choice for the mass: we also show results for a mass of the LSP equal to $200$ GeV in the appendix. The dark matter density is \cite{Seljak:2006bg} :
\begin{eqnarray}
0.106\ <\Omega\ h^2\ <\ 0.123 ,\\
\rm{with\ a\ central\ value\ around\ } 0.114 \nonumber
\label{matiere noire}
\end{eqnarray}

In the graph representing evolution of $Y_{lsp}$ as a function of the quantity $x=\frac{m_{lsp}}{T}$ there are two zones. The first zone is the equilibrium zone in which neutralinos are in thermal equilibrium and in which $Y$ decreases. The second zone is the zone after the freeze-out of neutralinos and in which $Y$ is constant. The freeze-out occurs at $x_f$. We call  $\Omega_{th}$ the thermal density of neutralinos. We find this approximate relation between $\Omega_{th}$ and $x_f$ :

\begin{eqnarray}
\Omega_{th}\ h^2 = 3.61\ 10^6 \ \frac{m_{lsp}}{1 \rm{GeV}}\ x_f^2\ e^{-x_f}
\label{omegath}
\end{eqnarray}
 
 To establish this relation we used \cite{Kolb:1990vq}. This corresponds to the relations below in which we have neglected the double log part. 
\begin{eqnarray}
x_{f}=\ln\lbrack 0.038(n+1)\frac{g}{g_\star^{1/2}}m_{pl}\ m_{lsp}\ \sigma_{0} \rbrack-(n+\frac{1}{2})\ \ln\lbrack\ln\lbrack 0.038(n+1)\frac{g}{g_\star^{1/2}}m_{pl}\ m_{lsp}\ \sigma_{0} \rbrack\rbrack \\          
\rm{and}\nonumber\\
\Omega_{th} \ h^2 = 1.07 \times 10^9 \frac{(n+1)x_{f}^{n+1}}{(g_{\star S} /g_{\star}^{1/2})m_{pl}\ \sigma_{0}},\\
\rm{with \ n=1 \ (for\ p-wave \ annihilation),\ g_{\star}=g_{\star S}=90,\ g=2} \nonumber 
\label{equa}
\end{eqnarray}  
where $g_{\star}$ is the number of effectively massless degrees of freedom, $ m_{pl}$ is the Planck mass.

\begin{figure}[h]
\centerline{\includegraphics{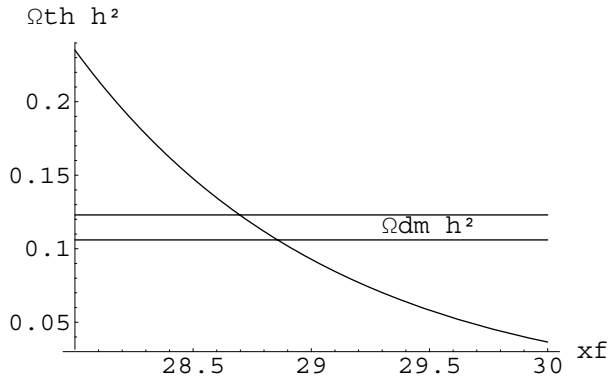}}
\caption{\label{omth}$\Omega_{th}\ h^2$ function of $x_f$. $\Omega_{dm}\ h^2$ is between the two straight lines.}
\end{figure}

We choose different values of $\Omega_{th} \ h^2$ and complete with the non thermal production coming from the gravitino decay. We call this non thermal production $\Delta \Omega \ h^2$. 

\begin{eqnarray}
0.106 \le \Omega_{th}\ h^2 + \Delta\Omega \ h^2 \le 0.123
\label{encadrement}
\end{eqnarray}

 As one gravitino produces one neutralino, we can write:

\begin{equation}
\Delta\Omega \ h^2 =\frac{ m_{lsp}\ s_0\ h^2}{\rho_c} \ \sum\limits_{k=0}^{k=\bar{n}} Y_{3/2}^k
\label{deltaomega}
\end{equation}

The index $\bar{n}$ corresponds to the last mode to be taken into account. It is the mode decaying just when the LSP decouples from the thermal bath. Only the gravitino modes decaying after the thermal decoupling of the neutralinos contribute to increase the quantity of neutralinos. 

\section{Model}
\subsection{Masses}
The thermal density $\Omega_{th}\ h^2$ and $x_f$ are related by the equation (\ref{omegath}). Knowing the thermal density we can calculate the non thermal density to respect the requirement (\ref{encadrement}). \\
Knowing $x_f$, we can calculate $T_f$ and then the mass of the last gravitino mode taken into account.\\
To do so we need a relation which allows us to convert temperature into time. In a radiation dominated Universe assuming that the entropy per comoving volume is conserved the relation between t and T is:
\begin{equation}
t=\frac{1}{2}\sqrt{\frac{90 \ M^2}{g_\star \pi^2}}\ T^{-2} \ {\rm GeV^{-1}}
\label{timetemperature}
\end{equation}
Where $M$ is the reduced Planck mass. 
This result (\ref{timetemperature}) is obtained by solving the Friedman equation in a radiation dominated Universe assuming constant entropy per comoving volume.\\ 
We use relation (\ref{timetemperature}) to calculate the mass of the mode $\bar{n}$ but also to calculate the mass of the 0 mode. For the mode $\bar{n}$, $g_{\star}=90$ and for the 0 mode, $g_{\star}=10$. We write equality between the life time of the gravitino mode (\ref{tempsdevie}) and the age of the universe\footnote{Even when the gravitino is not heavy enough to be non-relativistic when it is created, we can assimilate the life time of gravitino to the age of the Universe since most of its life is non-relativistic.} (\ref{timetemperature}) (after having converted (\ref{timetemperature}) into second) and we can calculate the mass of the corresponding gravitino.\\
 The gravitino which mainly decays at $T_f$ corresponds to the last mode to take into account in our Kaluza-Klein tower. We call it the mode $\bar{n}$.\\  
The mass of the zero mode is calculated the same way : we just put as a condition that it mainly decays at $T=1$ MeV not to disturb BBN.
For a temperature equal to 1 MeV we find with (\ref{timetemperature}) $t=0.76 \ s$. We then calculate $M_0$ with (\ref{tempsdevie}) by writing: $\tau_0=0.76 \ s$. We get $M_0=26.410\ \rm{TeV}$.\\
The mass of the mode $\bar{n}$ depends on the choice of $x_f$ which depends on the thermal density $\Omega_{th} \ h^2$. 

\subsection{Numerical results} 

We chose three different values of the density $\Omega_{th} \ h^2$.\\

The first chosen value corresponds to the central value $\Omega_{th}\ h^2=0.114$. This value is the value corresponding to our maximal thermal contribution. \\
The second value is calculated by choosing the maximum value for  $\sigma_0$. This cross section is generically bounded \cite{Kohri:2005ru} by: 
\begin{equation}
\sigma_{0} \leq \frac{\alpha}{m_{lsp^2}},   
\label{bound}
\end{equation}
With $\alpha \sim 10^{-2}$.
By using the maximum value for the annihilation cross section we find our minimal thermal contribution.\\
The last value is a medium value: we chose an $x_f$ value between our two $x_f$ extrema and calculate the thermal density corresponding to this $x_f$ value.\\    
Below is the table of values:\\

\begin{table}[h]

\begin{tabular}{|c|c|c|c|c|c|c|}

\hline
Cases
&$x_f$
&$T_f$(GeV)
&$\Omega_{th} \ h^2$
&$(\Delta\Omega_{th} \ h^2)_{min}$
& $(\Delta\Omega_{th} \ h^2)_{max}$
&$M_{\bar{n}}$(GeV)\\
\hline
Case 1
&28.78
&4.17 
&0.114
& -
&0.009
& $9.84 \times 10^6$ \\ 
\hline
Case 2
&29.60
&4.05 
&0.053
&0.053
&0.070
&$9.66 \times 10^6$ \\
\hline
Case 3 
&30.42
&3.94 
&0.025
&0.081
&0.098
&$9.48 \times 10^6$ \\
\hline
\end{tabular}
\caption{\label{troiscas}\textit{The three numerical cases for $m_{lsp}=120$ GeV}}
\end{table}

\subsubsection{Equations of constraints}

Using equation (\ref{deltaomega}) and replacing $Y_{3/2}^k$ by its value (\ref{abondancekk}), we obtain the non-thermal production of LSPs. \\
We have to distinguish two cases.\\
The first case is when the mass of the last mode to be taken into account (so the mode that we have called $\bar{n}$ and for which we have calculated the mass in the last section) is less than $T_R$ or equal to $T_R$ so $T_R \ge M^{\bar{n}}$. In that case if we look at the equations (\ref{abondancekk}) we deduce that $Y_{3/2}^k=Y_{3/2}^0$ and so we obtain from equation (\ref{deltaomega}):

\begin{eqnarray}
\Delta\Omega \ h^2=\frac{ m_{lsp} \ (\bar{n}+1) Y_{3/2}^0 \ s_0 \ h^2}{\rho_c} 
\label{domegan}
\end{eqnarray}

 We then should replace $\bar{n}$ by its value function of $R^{-1}$, $M^{\bar{n}}$ and $M^0$:  
\begin{equation} 
\bar{n}=\frac{M^{\bar{n}}-M^0}{R^{-1}}
\label{n}
\end{equation}
 Let us keep in mind that $\bar{n}$ is an integer number which means that $R^{-1}$ which is one of our variable must have a value which allows the ratio $\frac{M^{\bar{n}}-M^0}{R^{-1}}$ to be an integer.\\
So we will replace $\bar{n}$ by its value after having isolating it in one side of equation (\ref{domegan}). We define a function I which acts on real number and extracts the integer part.\\
We obtain:

\begin{eqnarray}
 & &R^{-1}=\left( M^{\bar{n}}-M^0 \right)\times
\nonumber\\
 & &\left( I\left[\Delta\Omega \ h^2 \frac{\rho_c}{m_{lsp} \ s_0 \ h^2}\ \frac{1}{1.9 \times 10^{-12}\times  \frac{T_{\rm R}}{10^{10}} \left[ 1 + 0.045 \ln \left( \frac{T_{\rm R}}{10^{10}}\right) \right]\left[ 1 - 0.028 \ln \left( \frac{T_{\rm R}}{10^{10}}\right) \right]}\right]-1 \right)^{-1}
\nonumber\\
\label{final}
\end{eqnarray}

In the above equation (\ref{final}) we use Table \ref{troiscas} to fix the limit values on $\Delta\Omega \ h^2$.\\
The above equation is valid for $T_R \ge M^{\bar{n}}$ and for a number of modes at least equal to 2. This last condition implies a limit on $T_R$: above the temperature corresponding to a number of modes equal to 2 there is only one mode which produces all the non thermal neutralinos.  We calculated those two limits on $T_R$ for the three cases. The results are presented with the graphs. \\
We also calculated the maximum reheating temperature allowed if there is only one mode for the three cases. The results are shown in Table \ref{MaxTR} and converge to the limit found in \cite{Ibe:2004tg} in the limit of vanishing thermal relic density. 

\begin{table}[h!]
\begin{center}
\begin{tabular}{|c|c|c|}
\hline
case 1
&case 2
&case 3\\
\hline
$1.45\ 10^9$ GeV
&$1.09\ 10^{10}$ GeV
&$1.52 \ 10^{10}$ GeV \\
\hline
\end{tabular}
\end{center}
\caption{\label{MaxTR}\textit{Maximum reheating temperature}}
\end{table}

If $T_R < M^{\bar{n}}$, it means that the mass of the last mode that should be taken into account has an abundance equal to zero using the rule given in (\ref{abondancekk}). In that case the last mass that it is effectively  taken into account is equal to $T_R$. We obtain the result for this case by just replacing $M^{\bar{n}}$ by $T_R$ in equation (\ref{final}).

\section{Gravitons}
As KK gravitinos are produced, there are also KK gravitons. We have to check that they do not disturb BBN since they are heavier than in the case of large extra dimensions usually considered. In our framework the usual astrophysical and cosmological constraints disappear like in the case of \cite{Kaloper:2000jb}. But there is a new problem coming from the fact that BBN should not be disturbed. The mass of the KK graviton is given by:

\begin{equation}
m_k=\frac{k}{R}
\end{equation}

So the mode 1 is the first massive mode with a mass equal to $\frac{1}{R}$. In the following we check that for $R^{-1}\geq 1$ TeV, BBN is not disturbed. For larger R the presence of KK gravitons starts affecting BBN as shown with our approximate method. A more accurate study is needed to check what is precisely the maximum size which can be allowed for the radius. This calculation is beyond the scope of the present work and will be discussed elsewhere.\\
The production equation of KK gravitons is given by \cite{Hannestad:2001nq} and \cite{Hall:1999mk}. We can use the same equation:

\begin{equation}
s \dot{Y_m}= \frac{11 m^5 T}{128 \pi^3 M^2}\ K_1(m/T),
\end{equation}

Where $K_1$ is the modified Bessel function of the first kind, m is the mass of the graviton, T the temperature and M the reduced Planck mass.\\

By integrating this equation between the reheating temperature and temperature T below 1 MeV, we find:

\begin{equation}
Y_m(T)=\frac{1485}{256\pi^5 M \ g^{\star 1/2}g^{\star}_S }\ m \int\limits_{m/T_R}^{\infty} x^3 K_1(x) dx
\label{gr}
\end{equation} 

where $g^{\star}$ and $g^{\star}_S$ are taken constant equal to 10 since most of the lifetime of the gravitons that we consider is after $T=1$ MeV.\\

We calculate the lifetime of KK gravitons with a mass above 1 TeV using the partial decay rates given in \cite{Han:1998sg}. We find :
\begin{equation}
\tau=3.310 \times \frac{\pi\ M^2}{m^3}\ {\rm GeV}^{-1}
\label{tdv}
\end{equation}

We then calculate the mass of the graviton which decays at $T=1$ MeV using equation (\ref{timetemperature}) and equation (\ref{tdv}). This graviton is the last one to be taken into account in our check. We find $m=37.4$ TeV and we calculate its lifetime $\tau=0.75$ s. As the radius is equal to 1 TeV the last mode that we consider in this study has a mass equal to 38 TeV.\\  
To check if KK gravitons disturb or not  BBN we use the curves given by Jedamzik in \cite{Jedamzik:2006xz} . Those curves give a limit on the density of a massive decaying particle (if it did not decay) as a function of its lifetime. We have calculated the density of gravitons at our epoch if they did not decay. We also have to know the hadronic branching ratio. We find $Bh=0.70$ for gravitons with mass above 1 TeV using the partial decay rates given in \cite{Han:1998sg}. \\
Those curves are made for one particle decaying and not like in the present case for a tower of particles. We can however roughly divide the study in two zones : one before $\tau=100$ s and the other after $\tau=100$ s.\\
In the first zone (i.e $\tau \ge 100$ s) we take as a limit this estimate value: $\Omega\ h^2=5\ 10^{-5}$ and before 100 s we take as a limit $\Omega\ h^2=10^{-1}$. \\
The mass of the graviton whose lifetime is 100 s is 7.33 TeV. So we sum the density contribution of the first seven modes of the first zone and we check that this sum is less than $5\ 10^{-5}$. We also sum the density contributions of the 31 modes of the second zone and check that this sum is less than $10^{-1}$. \\

The density for gravitons is :

\begin{equation}
\Omega \ h^2= \frac{\sum\limits_{k=1}^{k=n} m_k \ Y_k\  s_0\ h^2}{\rho_c}= \frac{s_0 \ h^2}{\rho_c} \sum\limits_{k=1}^{k=n} m_k^2 \frac{1485}{256\ \pi^5  M g^{\star 1/2}g^{\star}_S } \int\limits_{m_k/T_R}^{\infty} x^3 K_1(x) dx 
\end{equation}
Where we have used equation (\ref{gr}) in which we have replaced m by $m_k$ and $Y_m$ by $Y_k$ since each mass is associated to a mode.
We know that:
\begin{equation}
\int\limits_{m_k/T_R}^{\infty} x^3 K_1(x) dx \le 4.71
\end{equation}

This integral is equal to 4.71 if $m=0$ or if $m<<T_R$. If $m=T_R$, the integral is equal to 4.47. In the range of values for m and $T_R$ that we have, we can choose to set the value of the integral to 4.71: for masses below 7 TeV, it is obvious that $m/T_R <<1$ and for masses above 7 TeV and close to 38 TeV, the error that we make increases the density.   
      
This implies :

\begin{equation}
\Omega \ h^2 \approx 4.71 \frac{s_0 \ h^2 }{\rho_c} \sum\limits_{k=1}^{k=n}  \frac{1485\ m_k^2}{256\ \pi^5  M  g^{\star 1/2}g^{\star}_S }
\end{equation}

We can replace $m_k$ by $\frac{k}{R}$ and after doing the sum over k we find:
\begin{equation}
\Omega \ h^2 \approx 5.52 \ 10^{-14} \frac{ n (1+n)(1+2n)}{R^2}
\end{equation}

Note that $n=m_n R$ and therefore $\Omega \ h^2 $ is proportional to $R$. So the larger is the radius, the larger is $\Omega \ h^2 $.\\ 
The mass of the last mode of the first zone is 7 TeV so after having replaced n by $m_n R$ we can evaluate the contribution of the first seven modes. We find: 
\begin{equation}
\Omega \ h^2 = 4.64 \ 10^{-5}\ < \ 5 \ 10^{-5}
\end{equation}
 
The result is below the limit even if it is rather close to it. But this limit is also not completely defined for our particular study in which there is a tower of massive particles. We conclude as a first estimate that the density of the seven first modes of gravitons does not disturb BBN.\\

The mass of the last mode of the second zone is 38 TeV so after having replaced n by $m_n R$ we can evaluate the contribution of the 31 modes. In our calculation we take into account the seven first modes whose contribution is negligible. We find:

\begin{equation}
\Omega \ h^2= 6.30 \ 10^{-3}\ < \ 10^{-1}
\end{equation}

The result is well below the limit. We conclude as a first estimate that the density of the 31 modes of gravitons does not disturb BBN.\\

We can conclude this section by saying that the curves drawn from gravitinos constraints on dark matter density are valid for $R^{-1}\ge 1$ TeV. This new constraint comes from KK gravitons. A more precise study has to be done to check if $R^{-1}\ge 1$ TeV is a stringent limit or if it is possible to have a larger radius. 
 
\section{Results}

In this section we present our results for the three different cases. Let us remind that the first case corresponds to the case in which $\Omega_{th}\ h^2$ is our maximum $0.114$ and so the non-thermal part has to be less than $0.009$. The second case is our intermediate case in which $\Omega_{th} \ h^2=0.053$  and so the non-thermal part has to be more than $0.053$ and less than $0.070$. The third case corresponds to the minimum value for $\Omega_{th} \ h^2$ so $0.025$ and the non-thermal part has to be more than $0.081$ and less than $0.098$. The range of values that we have chosen for $T_R$ is from $10^5$ to $10^{10}$ GeV. All these cases are treated with $m_{lsp}=120$ GeV. We also treat three cases presented in the appendix for $m_{lsp}=200$ GeV.\\ 
The first general comment on the curves is that they are all increasing. The lower is the reheating temperature the higher is the number of modes needed (and so the smaller is $R^{-1}$) because the gravitino density (and so the non-thermal part of neutralino density) is nearly proportional to the reheating temperature.\\   
We can check that the curves of the first case are below the curves of the second case which are also below the curves of the third case: the more dark matter has to be produced the more modes is needed and the smaller $R^{-1}$ is. We can also check that all curves are converging to nearly the same $R^{-1}$: it is actually the maximum mass difference between the first mode which mainly decays at $T=1$ MeV and the last mode which mainly decays at $T_f$ and we have seen that our three different $M_{\bar{n}}$ have close values (see Table \ref{troiscas}). We notice in all the cases that a discrete structure is visible above a different temperature for the three cases. Actually the discrete structure is always present if we zoom enough on the graph. For high scale temperatures, this structure is always visible on our graphs since the number of modes is low.\\
We also trace on the figures the limit $R^{-1}\ge 1$ TeV to avoid perturbation of BBN by KK gravitons.\\               

\begin{figure}[htbp]
\centerline{\includegraphics[height=8cm]{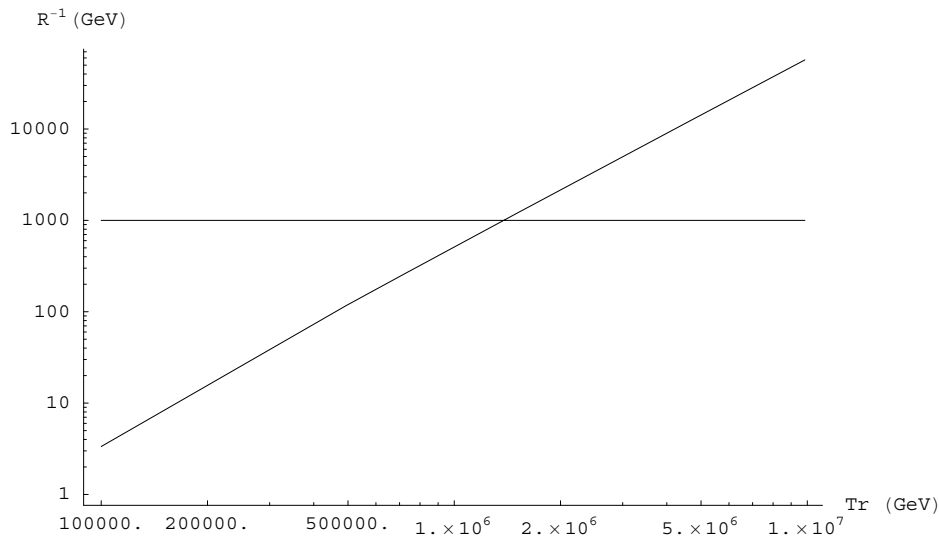}}
\caption{\label{1a}Case 1. $T_R$ less than $9.84 \ 10^6$ GeV. The excluded zone is below the diagonal curve and below the straight line $R^{-1}=1$ TeV if the KK gravitons constraint is taken into account.}
\end{figure}
The figure (\ref{1a}) is the case 1 for reheating temperatures below $9.84 \ 10^6$ GeV. This last temperature corresponds to the limit case $M^n=T_R$. The excluded zone is below the curve.\\
The maximum size for R corresponds to the minimum value for $R^{-1}$. Without taking into account KK gravitons constraint, the minimum would be $ R^{-1}=3.35$ GeV for a reheating temperature equal to $10^5$ GeV. So R should be equal or smaller than $5.89 \ 10^{-15}$ cm for $T_R=10^5$ GeV. Taking into account the limit $R^{-1} \ge 1$ TeV, R has to be equal or smaller than $1.97 \ 10^{-17}$ cm. 
For a reheating temperature of $9.84 \ 10^6$ GeV, R has to be equal or smaller than $3.46 \ 10^{-19}$ cm which corresponds to  $5.71 \ 10^4$ GeV for $R^{-1}$.\\ 
At the scale of the graph we cannot distinguish the discrete structure of the function. The maximum number of modes allowed should be equal to $21985$ for $T_R=10^5$ GeV without the constraint coming from KK gravitons and to $172$ for $T_R=9.84 \ 10^6$ GeV.\\
  
\begin{figure}[htbp]
\centerline{\includegraphics[height=8cm]{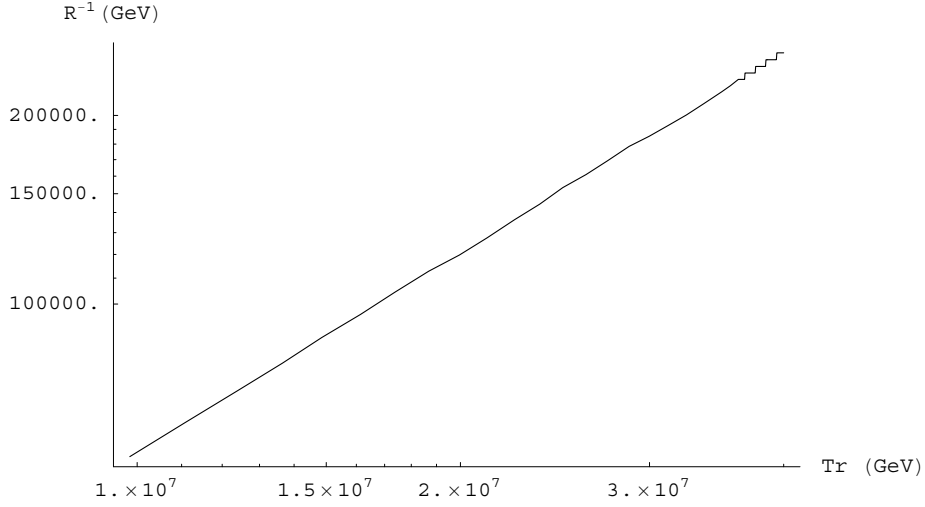}}
\caption{\label{1c}Case 1.  $9.84 \ 10^6 \ GeV \le T_R \le 4 \ 10^7 \ GeV$. The excluded zone is below the curve.}
\end{figure}
The figure (\ref{1c}) is the case 1 for reheating temperatures between $9.84 \ 10^6$ GeV  and $4 \ 10^7$ GeV. The excluded zone is below the curve. For a reheating temperature of $4 \ 10^7$ GeV, R has to be equal or smaller than $7.84 \ 10^{-20}$ cm (which corresponds to $2.52 \ 10^5$ GeV for $R^{-1}$). We distinguish a discrete structure on the graph. The maximum number of modes allowed for $T_R=4 \ 10^7$ GeV is 39.\\                                                                   

\begin{figure}[htbp]
\centerline{\includegraphics[height=8cm]{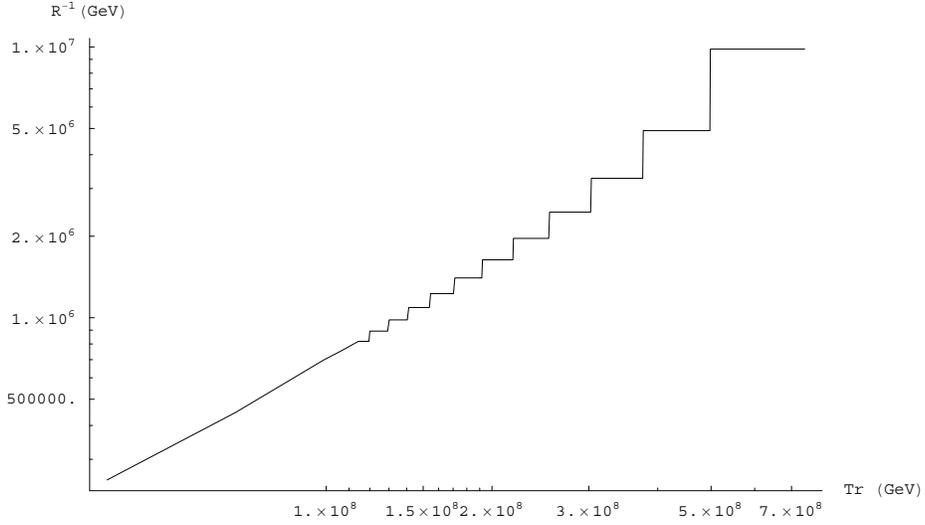}}
\caption{\label{1d}Case 1. $4\ 10^7 \ GeV \le T_R \le 7.396 \ 10^8 \ GeV $. The excluded zone is below the curve}
\end{figure}
The figure (\ref{1d}) is the case 1 for reheating temperatures between $4 \ 10^7$ GeV and $7.40 \ 10^8$ GeV. The excluded zone is below the curve. For a reheating temperature between $4.98 \ 10^8$ GeV and $7.40 \ 10^8$ GeV, the maximum number of modes allowed is 2 and R has to be equal or smaller than $2.01 \ 10^{-21}$ cm (which corresponds to $9.817 \ 10^6$ GeV for $R^{-1}$). Above the temperature $7.40 \ 10^8$ GeV, only a single mode is allowed and R has to be smaller than $2.01 \ 10^{-21}$ cm. The maximum allowed reheating temperature is given in Table \ref{MaxTR}:  $1.45 \ 10^9$ GeV. Above this temperature the number of neutralinos is above the observational limit (\ref{matiere noire}).\\         

\begin{figure}[htbp]
\centerline{\includegraphics[height=8cm]{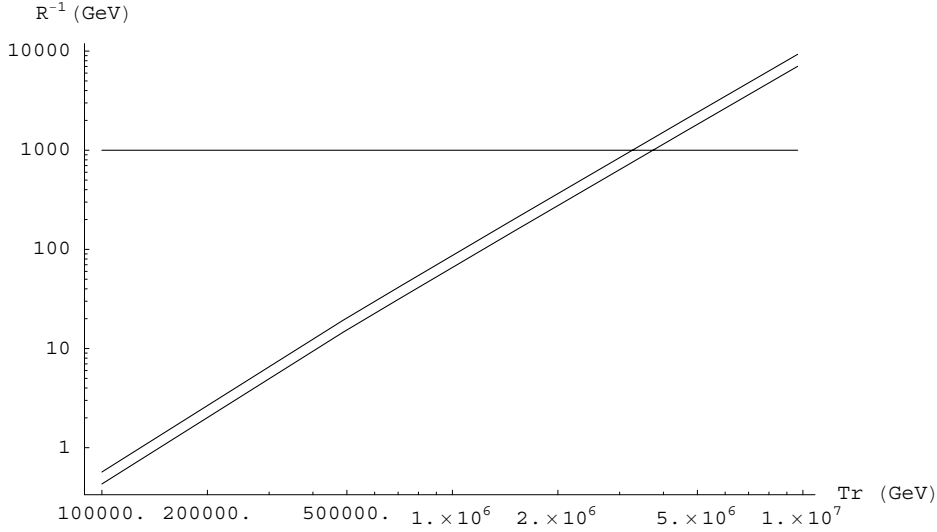}}
\caption{\label{2a}Case 2. $T_R$ less than $9.66 \ 10^6$ GeV. Only the band between the two diagonal curves is allowed. The zone below the straight line $R^{-1}=1$ TeV is excluded if KK gravitons constraint is taken into account.}
\end{figure}

The figure (\ref{2a}) is the case 2 for reheating temperatures below $9.66 \ 10^6$ GeV. Only the band between the two diagonal curves is allowed. It means that R has to be large enough to provide the necessary amount of dark matter but not larger than the value corresponding to the maximum amount of dark matter. If KK gravitons constraint is taken into account, the zone below $R^{-1}=1$ TeV is excluded.\\
 Without this constraint, $R^{-1}$ should be between $0.43$ GeV and $0.57$ GeV for $T_R= 10^5$ GeV: it means that R should be between $3.47 \ 10^{-14}$ cm and  $4.58 \ 10^{-14}$ cm. For $T_R=9.66 \  10^6$ GeV, $R^{-1}$ has to be between $7.02 \ 10^3$ GeV and $9.28 \ 10^3$ GeV, so R has to be between $2.13 \ 10^{-18}$ cm and $2.81 \ 10^{-18}$ cm.
For $T_R=10^5$ GeV, the number of modes should be between $129427$ and $170957$. For $T_R=9.66 \ 10^6$ GeV, the number of modes has to be between $1038$ and $1372$.\\
With KK gravitons constraint, the minimum value for $R^{-1}$ is 1 TeV so the maximum value for R is $1.97 \ 10^{-17}$ cm. It implies that there is a minimum value for $T_R$ corresponding to this value: the minimum value for $T_R$ is $3.26 \ 10^6$ GeV. Lower temperatures are excluded.\\

\begin{figure}[htbp]
\centerline{\includegraphics[height=8cm]{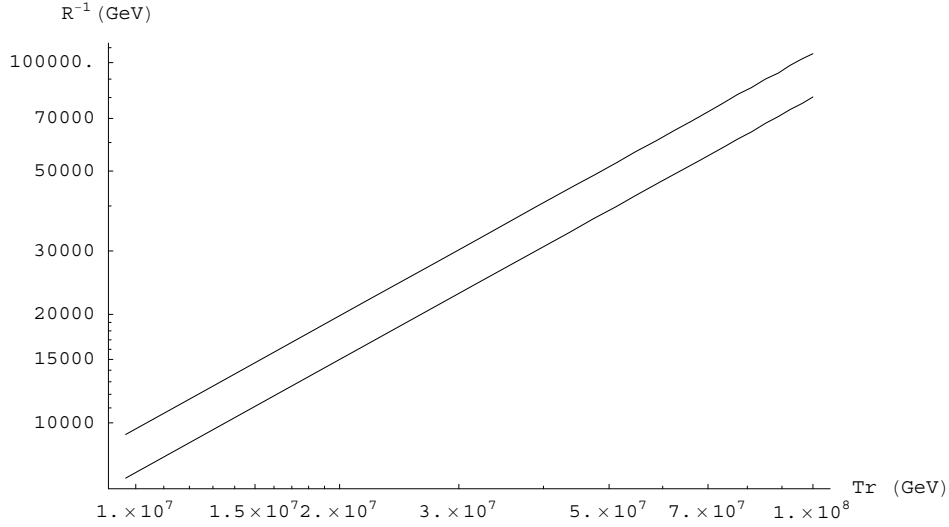}}
\caption{\label{2b}Case 2. $9.66 \ 10^6 \ GeV \le T_R \le  10^8 \ GeV $. Only the band between the two curves is allowed. }
\end{figure}
The figure (\ref{2b}) is the case 2 for reheating temperatures between $9.66 \ 10^6$ GeV and $10^8$ GeV. Only the band between the two curves is allowed. For $T_R=10^8$ GeV, $R^{-1}$ has to be between $8.03 \ 10^4$ GeV and $1.06 \ 10^5$ GeV so R has to be between $1.86 \ 10^{-19}$ cm and $2.46 \ 10^{-19}$ cm and the number of modes between $91$ and $120$ modes.\\    

\begin{figure}[htbp]
\centerline{\includegraphics[height=8cm]{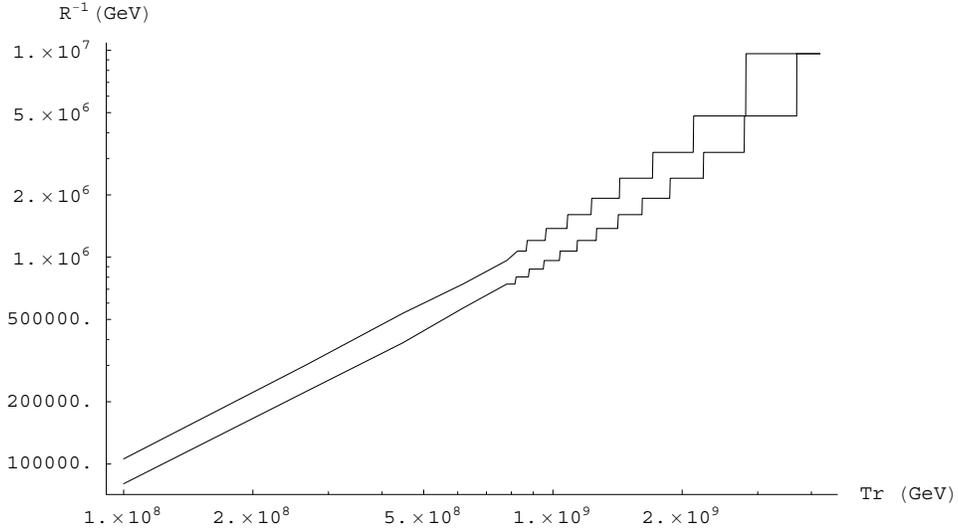}}
\caption{\label{2c}Case 2. $ 10^8 \ GeV \le T_R \le 4.190 \ 10^9  \ GeV $. Only the band between the two curves is allowed. } 
\end{figure}
The figure (\ref{2c}) is the case 2 for reheating temperatures between $10^8$ GeV and $4.19 \ 10^9$ GeV. Only the band between the two curves is allowed. For a reheating temperature between $3.70 \ 10^9$ GeV and $4.19 \ 10^9$ GeV, the number of modes has to be equal to 2 and R has to be equal to $2.05 \ 10^{-21}$ cm (which corresponds to $9.634 \ 10^6$ GeV for $R^{-1}$). Above the temperature $4.19 \ 10^9$ GeV , only a single mode is allowed and R has to be smaller than $2.05 \ 10^{-21}$ cm. The maximal allowed reheating temperature is given in Table \ref{MaxTR}:  $1.09 \ 10^{10}$ GeV. Above this temperature the density of neutralinos is above the observational limit given in (\ref{matiere noire}).\\ 

\begin{figure}[htbp]
\centerline{\includegraphics[height=8cm]{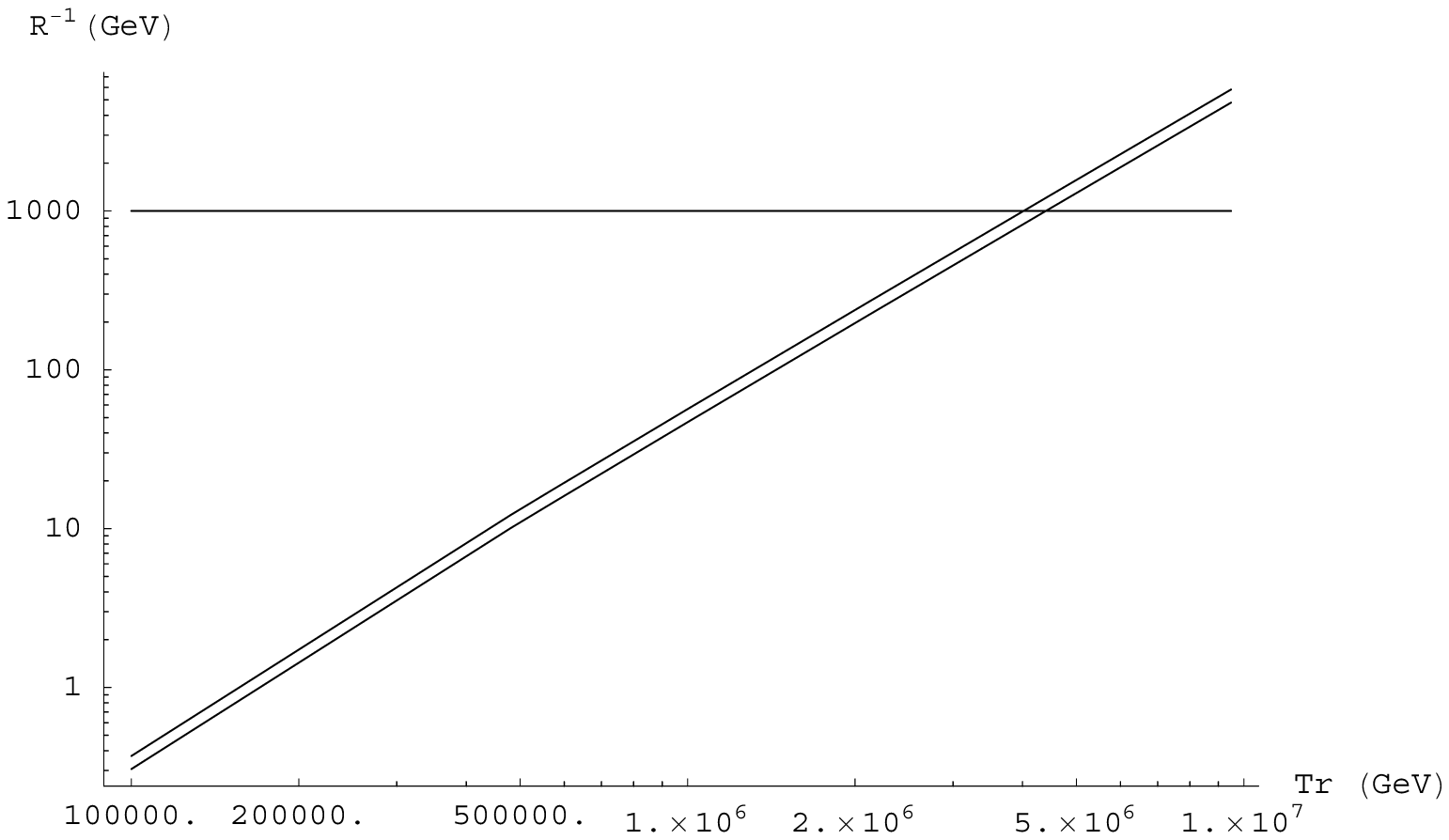}}
\caption{\label{3a}Case 3. $T_R$ less than $9.48 \ 10^6$ GeV. Only the band between the two diagonal curves is allowed. The zone below the straight line $R^{-1}=1$ TeV is excluded if KK gravitons constraint is taken into account.  }
\end{figure}
The figure (\ref{3a}) is the case 3 for reheating temperatures below $9.48 \ 10^6$ GeV. Only the band between the two diagonal curves is allowed as in case 2. If KK gravitons constraint is taken into account, the zone below $R^{-1}=1$ TeV is excluded. Without this constraint, $R^{-1}$ should be between $0.31$ GeV and $0.37$ GeV for $T_R=10^5$ GeV: it means that R should be between $5.33 \ 10^{-14}$ cm and $6.44 \ 10^{-14}$ cm. For $T_R=9.48 \  10^6$ GeV, $R^{-1}$ has to be between $4.81 \ 10^3$ GeV and $5.82 \ 10^3 $ GeV, so R has to be between $3.39 \ 10^{-18}$ and $4.10 \ 10^{-18}$ cm.
For $T_R=10^5$ GeV, the number of modes should be between $198804$ and $240333$. For $T_R=4.47 \ 10^6$ GeV, the number of modes has to be between $ 3568 $ and $4314  $.\\
With KK gravitons constraint, the minimum value for $R^{-1}$ is 1 TeV so the maximum value for R is $1.97 \ 10^{-17}$ cm. It implies that there is a minimum value for $T_R$ corresponding to this value: the minimum value for $T_R$ is $4.02 \ 10^6$ GeV. Lower temperatures are excluded.\\

\begin{figure}[htbp]
\centerline{\includegraphics[height=8cm]{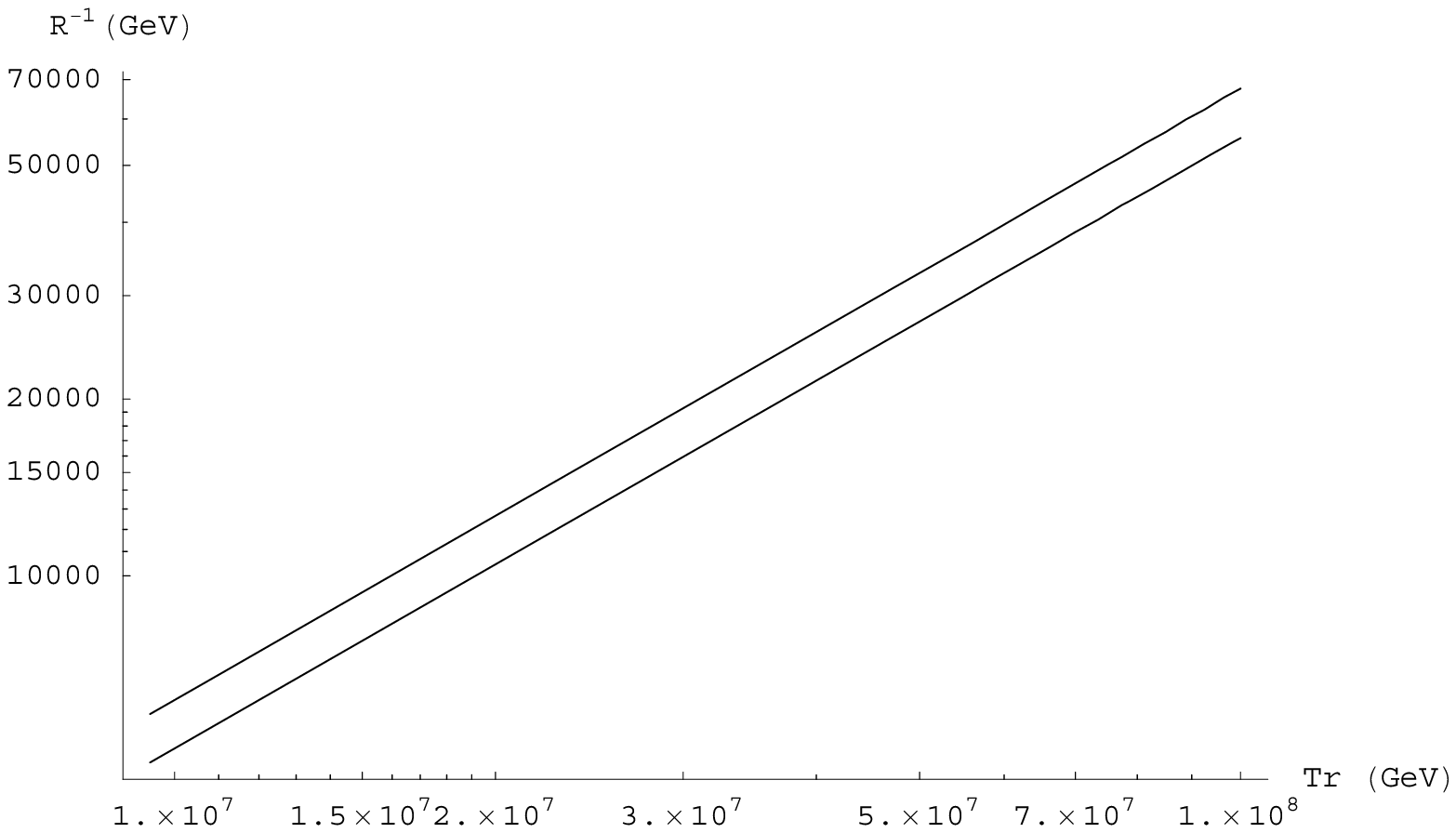}}
\caption{\label{3b}Case 3. $9.48 \ 10^6 \ GeV \le T_R \le  10^8 \ GeV $. Only the band between the two curves is allowed. }
\end{figure}
\begin{figure}[htbp]
\centerline{\includegraphics[height=8cm]{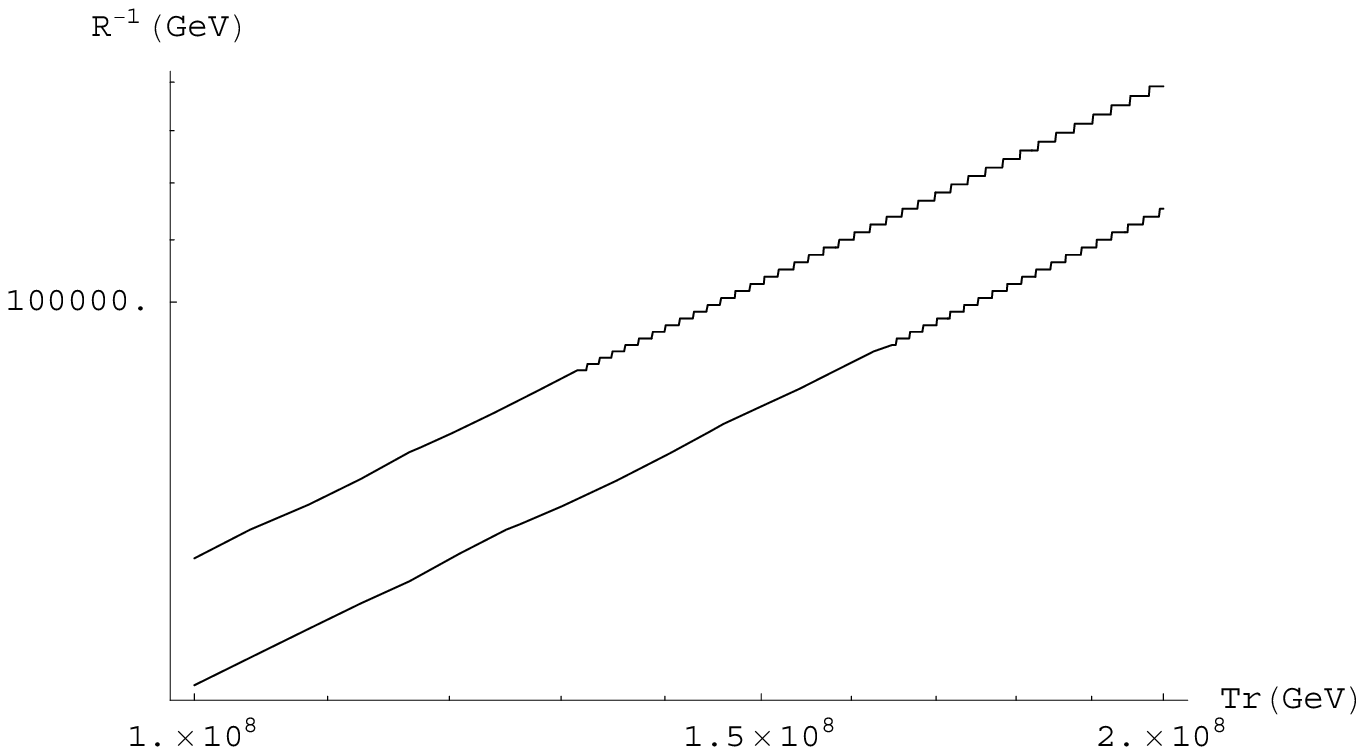}}
\caption{\label{3d}Case 3. $ 10^8 \ GeV \le T_R \le 2 \ 10^8  \ GeV $. Only the band between the two curves is allowed.}
\end{figure}
The figures (\ref{3b}) and (\ref{3d}) are the case 3 for reheating temperatures between $4.47 \ 10^6$ GeV and $10^8$ GeV for the figure (\ref{3b}) and between $10^8$ GeV and $2 \ 10^8$ GeV for the figure (\ref{3d}). Only the band between the curves is allowed. For $T_R=2 \ 10^8$ GeV, $R^{-1}$ has to be between $1.15 \ 10^5 $ GeV and $1.39 \ 10^{15} $ GeV so R has to be between $1.42 \ 10^{-19} $ cm and $1.71 \ 10^{-19}$ cm and the number of modes between $68 $ and $82 $ modes.\\

\begin{figure}[htbp]
\centerline{\includegraphics[height=8cm]{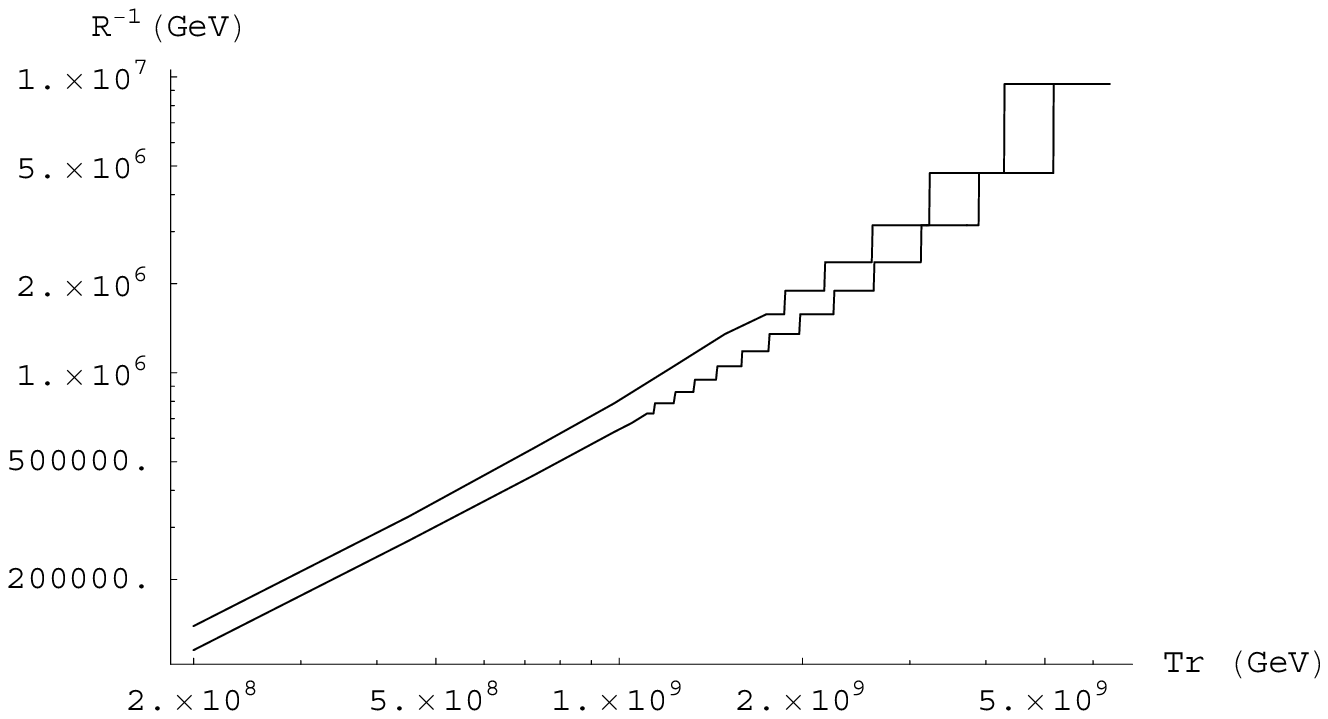}}
\caption{\label{3e}Case 3. $2 \ 10^8 \ GeV \le T_R \le 6.385 \ 10^9  \ GeV $. Only the band between the two curves is allowed.}
\end{figure}
The figure (\ref{3e}) is the case 3 for reheating temperatures between $2 \ 10^8$ GeV and $6.38 \  10^9$ GeV. Only the band between the two curves is allowed. For a reheating temperature between $5.17 \ 10^9$ GeV and $6.38 \ 10^9$ GeV, the number of modes has to be equal to 2 and R has to be equal to $2.09 \ 10^{-21}$ cm (which corresponds to $9.46 \ 10^6$ GeV for $R^{-1}$). Above the temperature $6.38 \ 10^9$ GeV, only a single mode is allowed and R has to be smaller than $2.09 \ 10^{-21}$ cm. The maximum allowed reheating temperature is given in Table \ref{MaxTR}:  $1.52 \ 10^{10}$ GeV. Above this temperature the density of neutralinos is above the observational limit given in (\ref{matiere noire}).\\

\newpage

 \section{Conclusion}

We have shown in this paper that for cosmological models with high reheating temperature i.e. $10^5 $ GeV to $10^{10}$ GeV - in the framework of a 5D supergravity compactified on $S^1/Z_2$ where matter and gauge fields live on tensionless branes at the orbifold fixed points - there are curves of constraints between the size R of the extra-dimension and the reheating temperature. It comes from the assumption that dark matter is made by the lightest supersymmetric particle which is supposed to be the lightest neutralino and that neutralinos density is a sum of a thermal production and a non thermal production from gravitinos decay. Gravitinos in the model do not disturb BBN because they are heavy enough to decay before BBN starts. Heavy gravitinos are natural in a certain class of Susy breaking models. For instance models which avoid problems from pure anomaly mediation and pure gravity mediation. The framework of the model described in this paper can be linked to Horava-Witten M-theory where a 5 dimensional stage of the universe appears in which bulk fields are gravitational and where supersymmetry is natural but also to theories of baryogenesis through leptogenesis which imply large reheating temperature.\\ 
The results that we obtain are independent from the susy mass spectrum since the gravitino is heavy enough to make negligible the influence of other susy particles. \\The results show that the size of the radius R is not only bounded by a maximum value but also by a minimum value in a wide range of possible values for the thermal production of neutralinos and in a wide range of values for the reheating temperature. To obtain a minimum size for the radius is a new result. \\
Kaluza-Klein modes of graviton are also present and may disturb BBN. We checked with an approximate method that for $R^{-1}$ above 1 TeV it is not the case. This already implies a bound on the reheating temperature which can not be lower than a minimum value in the cases where the radius R is bounded by a minimum value.\\


\begin{acknowledgments}

I thank very much Aldo Deandrea for his contribution to this work, Sacha Davidson and Karsten Jedamzik for useful discussions and comments and Guy Chanfray the Scientific Deputy Director of IPNL for his support.

\end{acknowledgments}

\newpage
\appendix*
\section{Numerical results and curves for $m_{lsp}=200$ GeV}

\begin{table}[h]

\begin{tabular}{|c|c|c|c|c|c|c|c|}

\hline
Cases
&$x_f$
&$T_f$(GeV)
&$\Omega_{th} \ h^2$
&$(\Delta\Omega_{th} \ h^2)_{min}$
& $(\Delta\Omega_{th} \ h^2)_{max}$
&$M_n$(GeV)
&Max.Reheat.Temp(GeV)\\
\hline
Case 1
&29.32
&6.82 
&0.114
& -
&0.009
& $1.37 \times 10^7$ 
& $8.84 \times 10^8$ \\
\hline
Case 2
&29.6194
&6.7523
&0.087
&0.019
&0.036
& $1.36 \times 10^7$ 
& $3.45 \times 10^9$           \\
\hline
Case 3 
&29.91
&6.69
&0.066
&0.040
&0.057
&$1.35 \times 10^6$
& $5.37 \times 10^9$   \\
\hline
\end{tabular}
\caption{\label{troiscas200}\textit{The three numerical cases for $m_{lsp}=200$ GeV}}
\end{table}

\begin{figure}[htbp]
\centerline{\includegraphics[height=8cm]{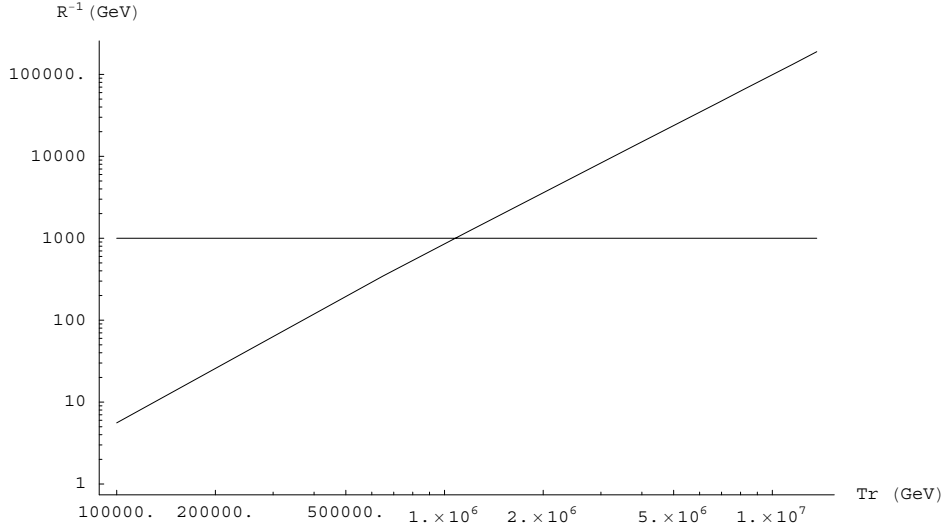}}
\caption{\label{2001bis}Case 1 for $m_{lsp}=200$ GeV. $T_R$ less than $ 1.37 \ 10^7  \ GeV $. The excluded zone is below the curve.}
\end{figure}

\begin{figure}[htbp]
\centerline{\includegraphics[height=8cm]{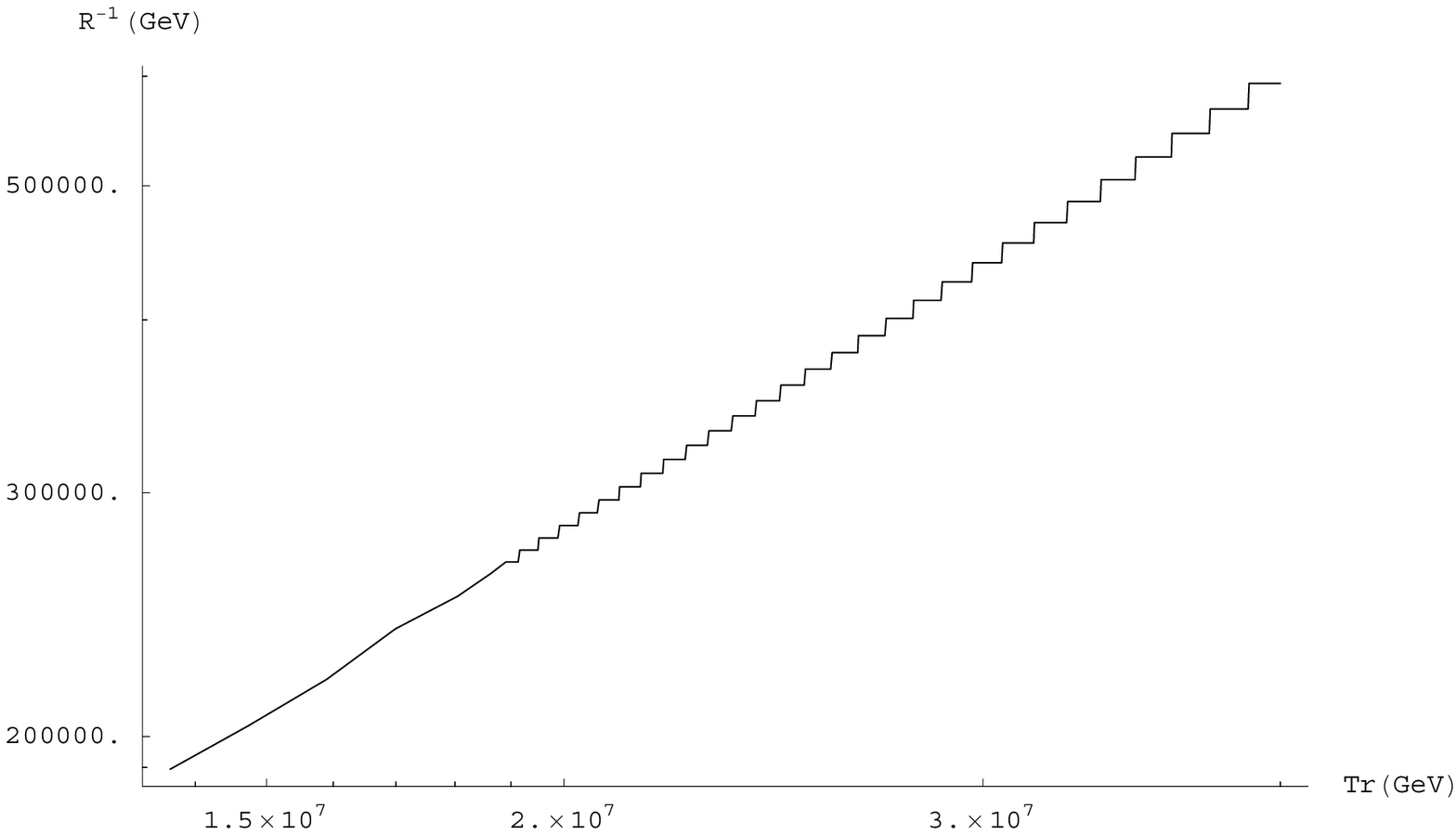}}
\caption{\label{2002}Case 1 for $m_{lsp}=200$ GeV. $ 1.37 \ 10^7 \ GeV \le T_R \le 4 \ 10^7  \ GeV $. The excluded zone is below the curve.}
\end{figure}

\begin{figure}[htbp]
\centerline{\includegraphics[height=8cm]{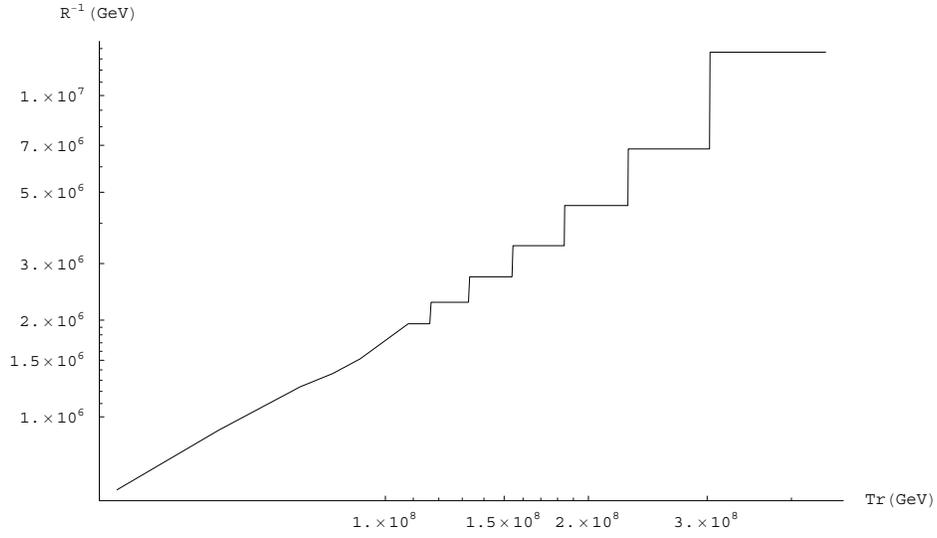}}
\caption{\label{2003}Case 1 for $m_{lsp}=200$ GeV. $ 4 \ 10^7 \ GeV \le T_R \le 4.49 \ 10^8  \ GeV $. The excluded zone is below the curve.}
\end{figure}

\begin{figure}[htbp]
\centerline{\includegraphics[height=8cm]{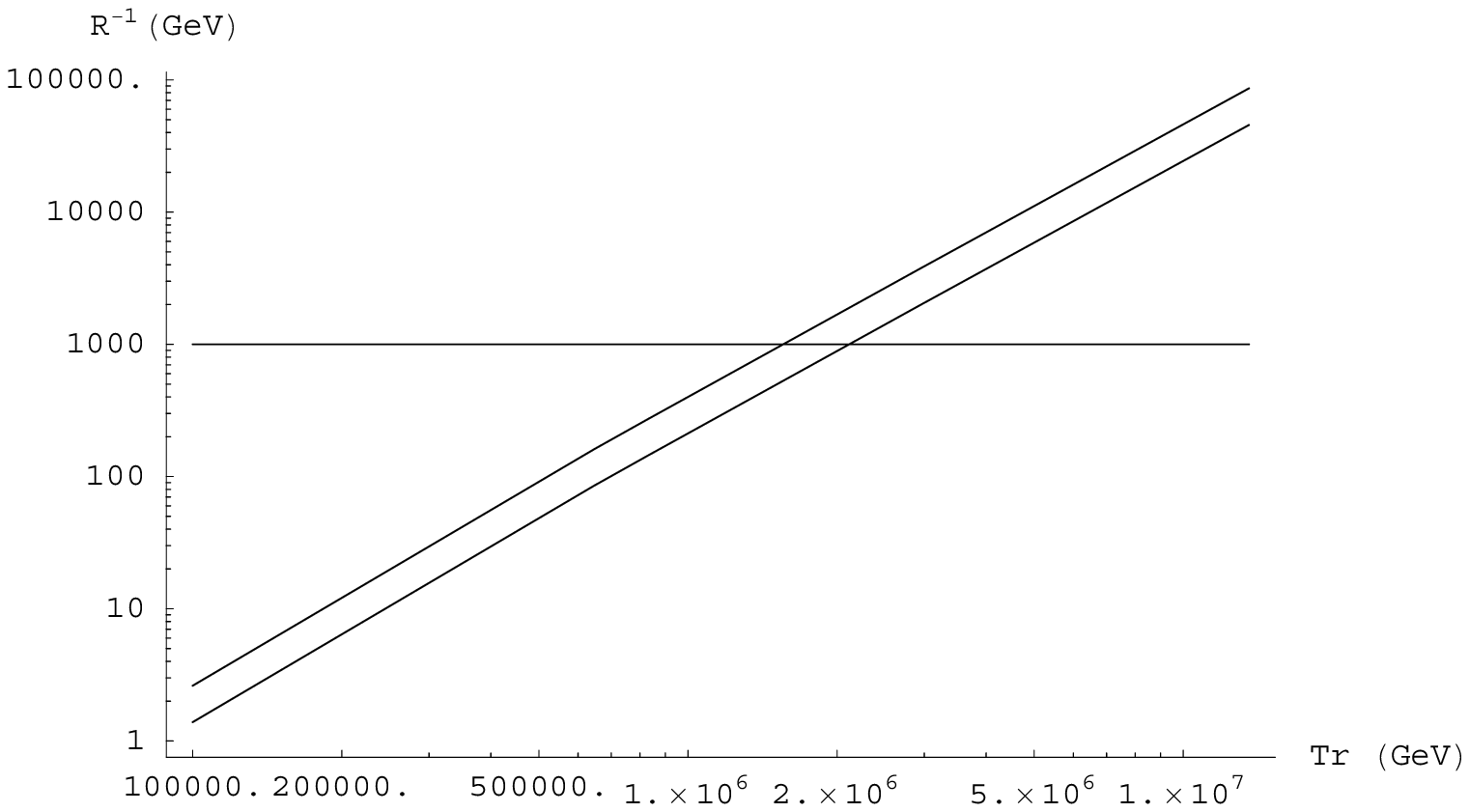}}
\caption{\label{2002a}Case 2 for $m_{lsp}=200$ GeV. $T_R$ less than $1.36 \ 10^7$ GeV. Only the band between the two diagonal curves is allowed. The zone below the straight line $R^{-1}=1$ TeV is excluded if KK gravitons constraint is taken into account.  }
\end{figure}

\begin{figure}[htbp]
\centerline{\includegraphics[height=8cm]{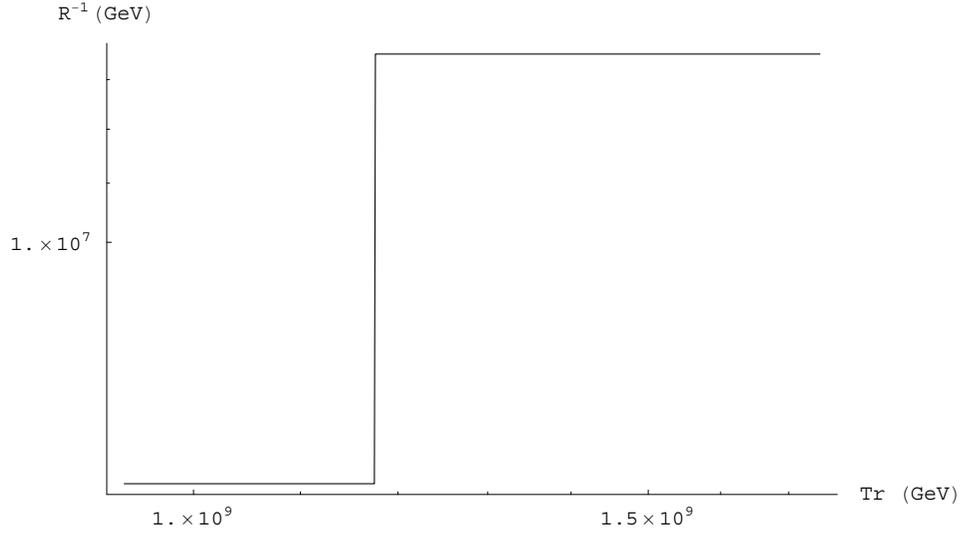}}
\caption{\label{2002d}Case 2 for $m_{lsp}=200$ GeV. $9.40 \ 10^8 \ GeV \le T_R \le  1.75\  10^9 \ GeV $ The zone below the curve is excluded. }
\end{figure}

\begin{figure}[htbp]
\centerline{\includegraphics[height=8cm]{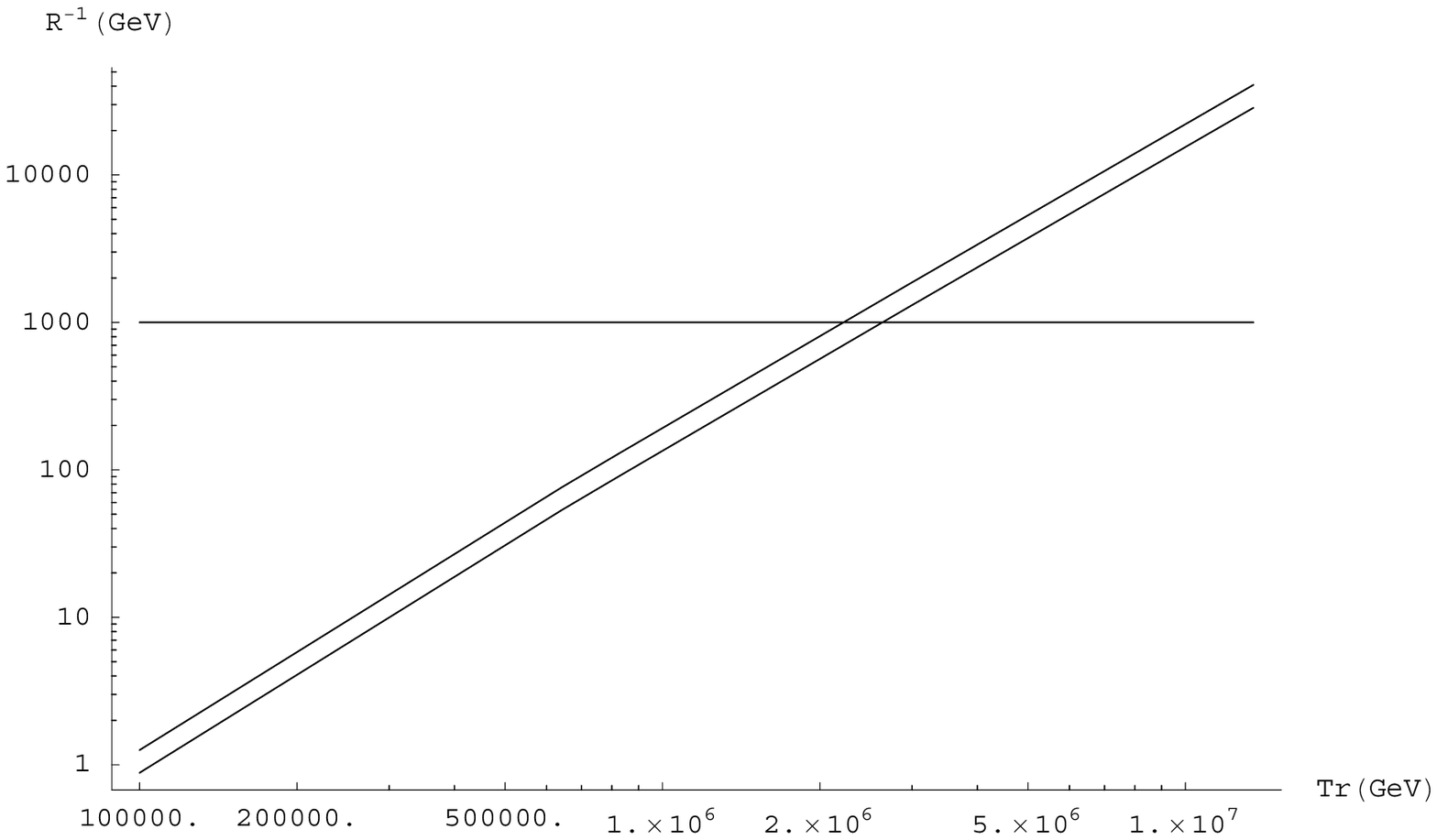}}
\caption{\label{2003a}Case 3 for $m_{lsp}=200$ GeV. $T_R$ less than $1.35 \ 10^7$ GeV. Only the band between the two diagonal curves is allowed. The zone below the straight line $R^{-1}=1$ TeV is excluded if KK gravitons constraint is taken into account.  }
\end{figure}

\begin{figure}[htbp]
\centerline{\includegraphics[height=8cm]{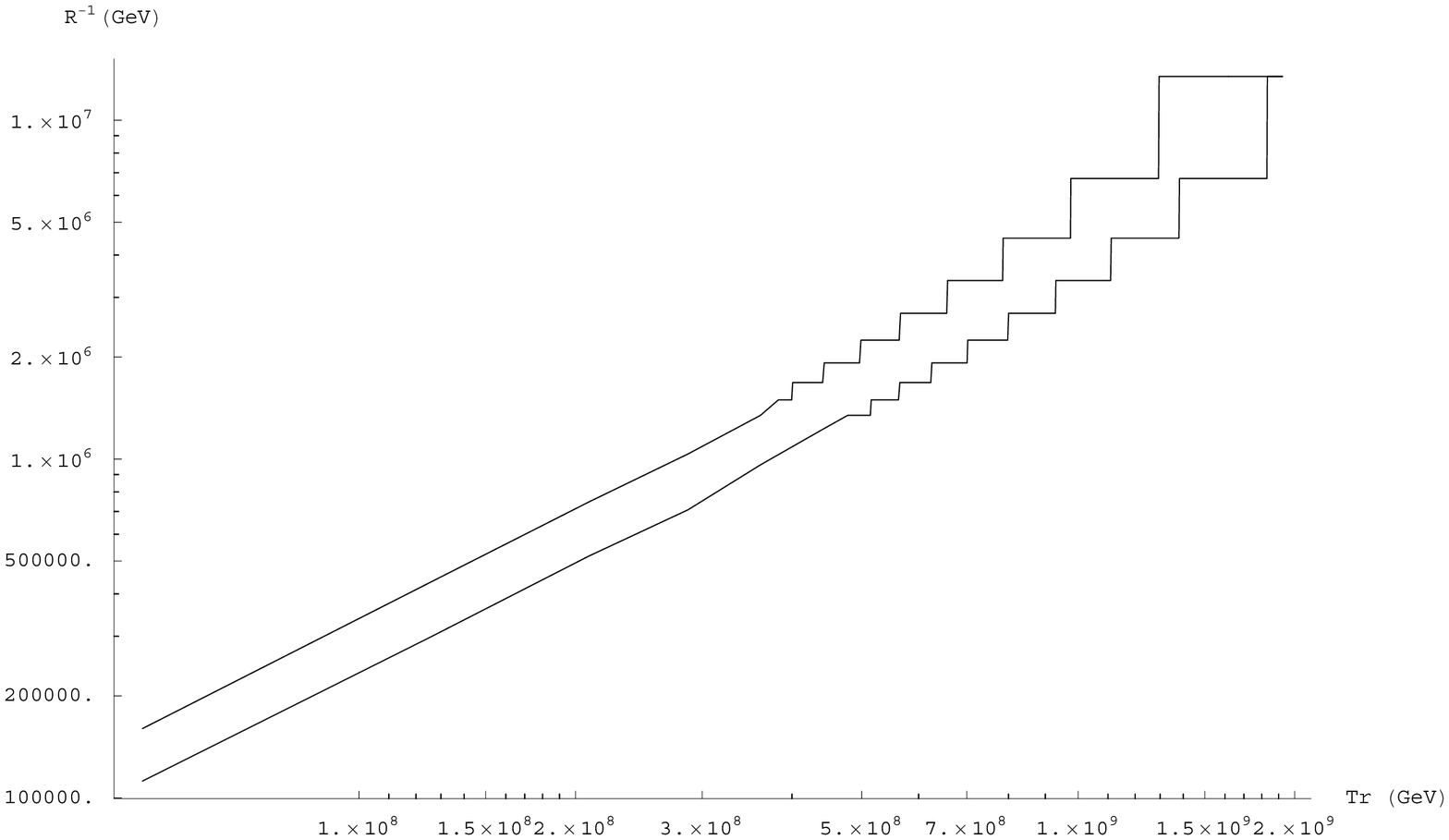}}
\caption{\label{2003c}Case 3 for $m_{lsp}=200$ GeV. $5 \ 10^7 \ GeV \le T_R \le  1.92\  10^9 \ GeV $ Only the band between the two curves is allowed. }
\end{figure}

The figures (\ref{2001bis}), (\ref{2002}) and (\ref{2003}) represent the case 1 but for $m_{lsp}=200$ GeV instead of $120$ GeV.  We notice that the curves of that case are above the curves for the case 1 with $m_{lsp}=120$ GeV. We can explain this by looking at equation (\ref{final}): we notice that $R^{-1}$ is nearly proportional to $m_{lsp}$. Physically it means that if the elementary energy of one LSP increases, less number density of LSPs is required and so $R^{-1}$ can be bigger. $M_{\bar{n}}$, $T_f$ and the maximum reheating temperature are slightly different from the case with $m_{lsp}=120$ GeV (see Table \ref{troiscas200}).\\ 
The figures of case 2 (fig.\ref{2002a} to fig.\ref{2002d}) and case 3 (fig.\ref{2003a} to fig.\ref{2003c}) for $m_{lsp}=200$ GeV have the same general behavior as cases 2 and 3 for $m_{lsp}=120$ GeV: the curves are just above the ones for $m_{lsp}=120$ GeV.\\  
For the case 2 the minimum reheating temperature is $1.55 \ 10^6$ GeV if the constraint on KK gravitons is taken into account (see fig.\ref{2002a}). For the case 3 the minimum reheating temperature is $2.216 \ 10^6$ GeV if the constraint on KK gravitons is taken into account (see fig.\ref{2003a}).\\ 

\begin{figure}[htbp]
\centerline{\includegraphics[height=8cm]{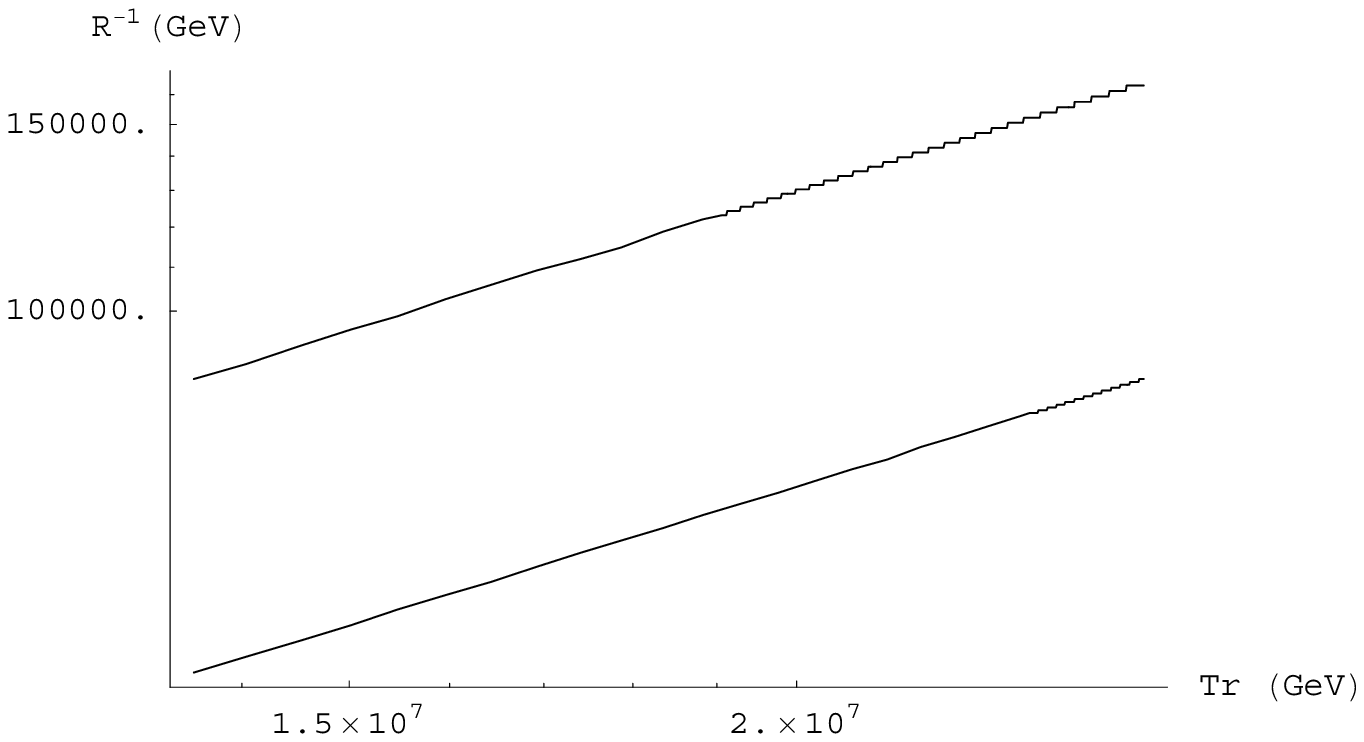}}
\caption{\label{2002b}Case 2 for $m_{lsp}=200$ GeV. $1.36 \ 10^7 \ GeV \le T_R \le  2.5\  10^7 \ GeV $ Only the band between the two curves is allowed. }
\end{figure}

\begin{figure}[htbp]
\centerline{\includegraphics[height=8cm]{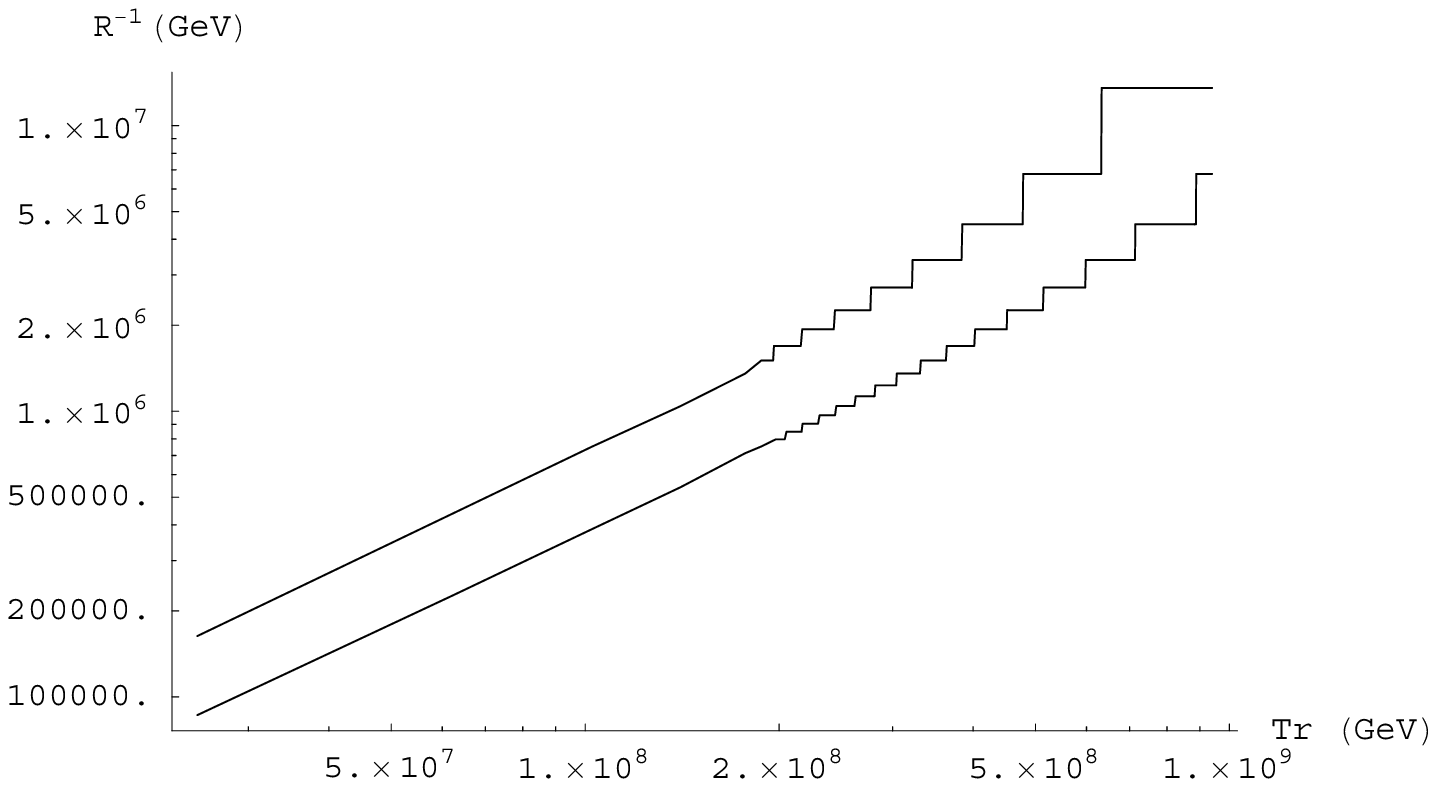}}
\caption{\label{2002c}Case 2 for $m_{lsp}=200$ GeV. $2.5 \ 10^7 \ GeV \le T_R \le  9.40\  10^8 \ GeV $ Only the band between the two curves is allowed. }
\end{figure}

\begin{figure}[htbp]
\centerline{\includegraphics[height=8cm]{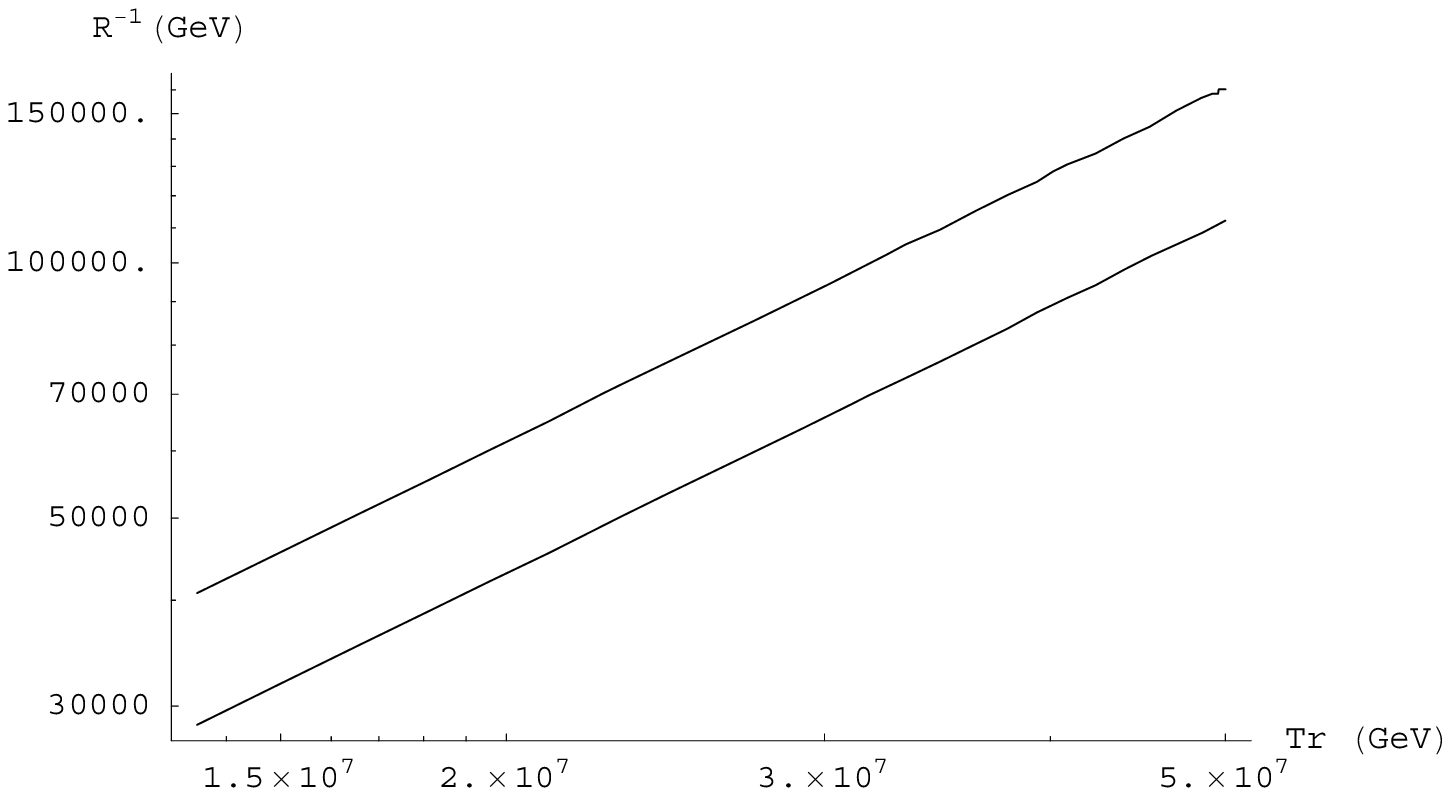}}
\caption{\label{2003b}Case 3 for $m_{lsp}=200$ GeV. $1.35 \ 10^7 \ GeV \le T_R \le  5\  10^7 \ GeV $ Only the band between the two curves is allowed. }
\end{figure}

\end{document}